\documentclass[journal,final]{IEEEtran}
\IEEEoverridecommandlockouts

\usepackage{graphicx,psfrag}
\usepackage{amsmath,amssymb,amsfonts,mathrsfs,mathtools}
\usepackage{color,epstopdf}
\usepackage{algorithmic}
\usepackage{enumerate} 
\usepackage{subfigure}
\usepackage{floatrow}
\usepackage{float}
\usepackage{stfloats}
\usepackage{flushend} 
\usepackage{bbold}
\usepackage{dsfont}
\usepackage{mathrsfs}
\usepackage{relsize}
\usepackage{relsize}

\usepackage{mathabx}

\usepackage[noadjust]{cite}
\newcommand{\R}{\ensuremath{\mathbb{R}}}
\newcommand{\N}{\ensuremath{\mathbb{N}}}

\newtheorem{example}{Example}
\newtheorem{rem}{Remark}
\newtheorem{assumption}{Assumption}

\newtheorem{theorem}{Theorem}
\newtheorem{lem}{Lemma}
\newtheorem{proposition}{Proposition}

\begin{document}

\title{Event-triggered Control From Data}

\author{C. De Persis, R. Postoyan, and P. Tesi 
\thanks{C. De Persis is with ENTEG and the J. C. Willems Center for
Systems and Control, 
University of Groningen, 9747 AG Groningen, The Netherlands.
Email: {\tt\small c.de.persis@rug.nl}. \\
R. Postoyan is with the Université de Lorraine, CNRS, CRAN, 
F-54000 Nancy, France. His work is supported by ANR via grant HANDY, number ANR-18-CE40-0010. E-mail: {\tt\small romain.postoyan@univ-lorraine.fr}.\\
P. Tesi is with DINFO, University of Florence, 50139 Firenze, Italy. E-mail: {\tt\small pietro.tesi@unifi.it}.}
}

\maketitle
\begin{abstract}
We present a data-based approach to design event-triggered state-feedback controllers for unknown continuous-time linear systems affected by disturbances. By an event, we mean state measurements transmission from the sensors to the controller over a digital network. By exploiting a sufficiently rich finite set of noisy state measurements and inputs collected off-line, we first design a data-driven state-feedback controller to ensure an input-to-state stability property for the closed-loop system ignoring the network. We then take into account sampling induced by the network  and we  present robust data-driven triggering strategies to (approximately) preserve this stability property. The approach is general in the sense that it allows deriving data-based versions of various popular triggering rules of the literature. 
In all cases, the designed transmission policies ensure the existence of a (global) strictly positive minimum inter-event times thereby excluding Zeno phenomenon despite disturbances.  These results can be viewed as a step towards 
plug-and-play control for networked control systems, i.e., mechanisms
that automatically learn to control and to communicate over a network.  
\end{abstract}

\section{Introduction}\label{sec:intro}

\IEEEPARstart{E}vent-triggered control is an implementation paradigm, which consists in transmitting data between the plant and its controller whenever a state- or output-dependent criterion is verified. The underlying idea is to generate communications between the plant and its controller only when this is needed to achieve the desired control objectives, as opposed to classical time-triggered (periodic) strategies for which the communication instants depend on the elapsed time and not on the actual system needs. Event-triggered control is motivated by resource aware scenarios where communicating, computing or updating the control input comes with a certain cost, such as networked control systems and embedded systems. Since the results in \cite{Arzen-99,Astrom-Bernhardsson-02} and the pioneering work in \cite{Tabuada07}, various event-triggered control strategies have emerged in the literature for a range of set-ups and control problems, see, e.g., \cite{Heemels2012,Postoyan2014}.

The vast majority of the literature on event-triggered control focuses on model-based approaches, in the sense that both the feedback law and the triggering condition are designed by relying on a model of the plant dynamics. These results are therefore not applicable when first-principle models are not conceivable, or when exact/accurate enough system identification is impossible because of noisy data. In this case, an alternative consists in designing the controller directly based on available input-state/output data, we talk of data-driven control, see e.g.,  
 \cite{safo1995,HGGL98,CLS02,fliess-join-ijc2013,sznaier2021}
for earlier contributions and one recent survey on the topic. 
This paradigm is also appealing as it may ease the controller design step.
Few techniques are currently available in the literature to design data-driven event-based controllers see, e.g., \cite{liu-et-al2022data,cordovil2022learning,Digge-pasumarthy-ecc22,berbDTb2022}, which consider discrete-time systems, and the recent work in \cite{qi-et-al-tie2022} dedicated to continuous-time systems. In \cite{liu-et-al2022data,cordovil2022learning,Digge-pasumarthy-ecc22}, no disturbances act on the data collected off-line, which simplifies the learning, in particular, for linear time-invariant systems,  an exact data-based representation can be obtained in this case under mild conditions \cite{willems2005}. This is not the case of  \cite{berbDTb2022} where the controller and the triggering policy are designed 
once and for all with one single batch of \emph{noisy} data collected from the system. The idea is that, under certain conditions, this data batch  returns a (non-parametric) system model.  In turn, this makes it possible to cast control and triggering rule design
as data-dependent problems,  in particular via data-based linear matrix inequalities (LMI). The authors of \cite{berbDTb2022} then derive a dynamic event-triggered control policy inspired by the model-based technique in \cite{Girard-tac15}. However, while off-line data are noisy, disturbances are ignored when implementing the controller. In \cite{qi-et-al-tie2022}, on the other hand, disturbances are considered both for learning and during closed-loop operation and a specific dynamic triggering strategy is presented to ensure an $\mathcal{L}_2$-stability property.

There is therefore a strong need for general data-based event-triggered control techniques that are applicable to continuous-time systems, as many real-life processes have a continuous-time nature, and that are robust to disturbances affecting \emph{both} the data acquisition phase and the closed-loop operations. 
In this context, we present a data-based  approach to design state-feedback event-triggered controllers for unknown stabilizable continuous-time linear  systems affected by  disturbances. Like in \cite{qi-et-al-tie2022}, noisy data is explicitly accounted both off-line, when acquiring data, and on-line, when implementing the designed controller. Noisy data in the acquisition phase prevent  exact system identification, and this calls for control design routines  that are robust to uncertainty. Noise during the closed-loop operations, on the other hand, is 
even more problematic because it may lead to Zeno behavior even with exact modelling \cite{Borgers-Heemels-tac14}, meaning that an infinite number of transmissions occur in finite time. To address these challenges, we proceed by emulation and, as a first step, we design a state-feedback controller that stabilizes the closed-loop system in absence of sampling, in the sense that an input-to-state stability (ISS) property holds. We then take into account sampling induced by the network and we  design data-driven triggering techniques based on the relative threshold technique of \cite{Tabuada07} that preserves (approximately) the original stability property of the closed-loop system. In particular, we first develop data-driven mixed absolute/relative threshold event-based policies \cite{Borgers-Heemels-tac14}. In this case, the closed-loop system is guaranteed to satisfy a practical ISS property, where the adjustable parameter is the absolute threshold. Moreover, we prove the existence of a strictly positive uniform minimum inter-event times between any two transmissions and we provide a data-based estimate of it. 

To preserve the asymptotic nature of the stability property of the continuous-time closed-loop system in absence of sampling, we  present an alternative triggering rule based on time-regularization of the relative threshold, see e.g.,  \cite{dpt_DOS_2015,Abdelrahim2017,Dolk-et-al-tac17}. By doing so, we enforce a given strictly positive minimum inter-event time after which a relative threshold condition is checked thereby excluding Zeno phenomenon by design. The maximum value of this enforced minimum time between any two transmissions is given by an expression, which depends on the data collected off-line only. While the time-regularized triggering condition ensures a stronger stability property compared to the mixed strategy, it may generate more transmissions as suggested by a numerical example we provide. 
From a technical viewpoint, the proposed design conditions take the form of semi-definite programs (SDP), in particular feasibility of optimization problems with LMI constraints. This is appealing because many results on event-triggered control developed in a model-based context involve LMI formulations, see 
\cite{Heemelsb2012,Abdelrahim2017}. 

Afterwards, we illustrate the generality of the approach by 
designing 
robust data-based versions of several major triggering rules of the literature originally developed in a model-based setting, namely: (i) quadratic triggering policies, e.g., \cite{Heemels2012}; (ii) dynamic event-triggered control \cite{Girard-tac15,Dolk-et-al-tac17} where we consider a more general form compared to \cite{qi-et-al-tie2022} (see Remark \ref{rem:qi-et-al} for more details);  and (iii)  decreasing threshold on a Lyapunov function designed ignoring sampling, like in  \cite{Wang-Lemmon-aut11,Mazo-Anta-Tabuada-Aut10}. In all cases, ISS properties are established and a global (i.e., independent of the initial conditions of the system) minimum inter-event time is guaranteed to exist despite disturbances acting on the system. In addition, data-based estimates of these minimum inter-event times are provided. These results can be viewed as a step towards plug-and-play control for networked control systems, i.e., mechanisms
that automatically learn to control and to communicate over a network.

The remainder of this paper is as follows. We introduce the framework of interest in Section \ref{sec:framework}. We then first concentrate on the noise-free case to put in place the required background and techniques in Section  \ref{sec:exact}. The main results dedicated to noisy systems are provided in Section \ref{sec:noise}, where we establish (practical) ISS properties as well as the existence of a global minimum inter-event time. The extension to other triggering policies is presented  in Section \ref{sect:other-triggering-rules}. 
Section \ref{sec:conc} ends the paper with concluding remarks, and the appendix contains a couple of additional technical results.
Throughout the paper we give numerical examples to substantiate the analysis.




{\it Notation.}  $\mathbb N_0:=\{0,1,2,\ldots\}$ denotes the set of non-negative integers and $\N:=\N_0\backslash\{0\}$. $\R$ stands for the set of real numbers, $\R_{\geq 0}:=[0,\infty)$ and $\R_{>0}:=(0,\infty)$. Given a symmetric matrix $M$, the notation
$M \succ 0$ ($M \succeq 0$) and $M \prec 0$ ($M \preceq 0$)
means that $M$ is positive and, respectively, negative (semi)-definite. 
Given a matrix $M$, $M^\top$ denotes the transpose of $M$. The notation $I$ and $\mathbf{0}$ stand for the identity matrix and the zero matrix, respectively, whose dimensions depend on the context. For the sake of convenience, we write $\left[\begin{smallmatrix}
A & B \\ \star & C
\end{smallmatrix}\right]$ for $\left[\begin{smallmatrix}
A & B \\ B^\top & C
\end{smallmatrix}\right]$ where $A,B,C$ are matrices of appropriate dimensions. The induced $2$-norm of a matrix is denoted $\|\cdot\|$. 
Given $x\in\R^{n}$ and $y\in\R^{m}$ with $n,m\in\N$, 
we use the notation $(x,y)$ to denote $(x^\top,y^\top)^\top$. 
Given $x,y\in\R^{n}$ with $n\in\N$, $\left\langle x,y \right\rangle = x^\top y$ is the dot product. For $d:\R_{\geq 0}\to\R^{n}$ with $n\in\N$, the  supremum of $d$ on $[0,t]$ with $t\geq 0$ is $\|d\|_{[0,t]}:=\sup_{s\in[0,t] } |d(s)|$, and  $\|d\|_{\infty}:=\sup_{t\geq 0} |d(t)|$,  where $ |\cdot |$ denotes the Euclidean norm. 

\section{Framework} \label{sec:framework}

Consider the continuous-time linear system
\begin{equation}
\label{system}
\dot x(t) = A x(t) + B u(t) + d(t),
\end{equation}  
where $x(t) \in \mathbb R^{n}$ is the state and $u(t) \in \mathbb R^{m}$ is the control input at time $t\in\R_{\geq 0}$ with $n,m\in \N$. Input disturbance $d \in \R_{\geq 0}\to \mathbb R^{n}$ is unknown, Lebesgue measurable and bounded in the sense that $ \|d\|_{\infty} < + \infty$. Matrices
$A$ and $B$ are real, constant, \emph{unknown} and assumed to be such that $(A,B)$ is stabilizable.  We will not need to directly
check stabilizability of $(A,B)$ in the sense that this property is necessary for the feasibility of the design data-based programs that we will consider.

We investigate the scenario where plant \eqref{system} is connected to its controller via a network. In particular, state measurements are sent from the sensors to the controller via the digital channel, and the controller is directly connected to the actuators; although the results presented later in Section \ref{sect:other-triggering-rules} also allow considering the case where the network is instead between the controller and the actuators. 
Our objective is to design a state-feedback event-triggered controller to stabilize, in a sense made precise in the sequel,  system (\ref{system}). In particular, we aim at designing  a linear state-feedback law with gain $K\in\R^{m\times n}$  and a triggering policy that defines the sequence of transmission (or sampling) instants $\{t_k\}_{k \in \mathcal{N}}$ with $\mathcal{N}\subseteq\mathbb N_0$ for each solution to the system. Without loss of generality, we consider that a transmission occurs at $t=0$ so that $t_0=0$. We implement the controller using zero-order-hold devices, which leads to the control input
\begin{eqnarray} \label{control}
u(t) = Kx(t_k), \quad t \in [t_k,t_{k+1}). 
\end{eqnarray}
The solutions to (\ref{system}) in closed-loop with (\ref{control}) are understood in the Carath\'eodory sense\footnote{At the exception of the second part of Section \ref{subsect:dynamic-etc} as specified there.}, i.e., the solution flows on $[t_k,t_{k+1}]$ and experiences a jump at $t_{k+1}$ for $k,k+1\in\mathcal{N}$. We will establish later in the paper that, for any $k\in\mathcal{N}$, $t_{k+1}-t_{k}$ is  lower bounded by a strictly positive constant independent of $k$ and of the initial condition so that the $t_k$'s do not accumulate (i.e., Zeno phenomenon does not occur). Also, by a solution, we always mean a maximal solution, i.e., one that cannot be extended.

Various solutions to this problem are available in the literature when $A$ and $B$ are known see, e.g., \cite{Tabuada07,Girard-tac15,Postoyan2014,Heemels2012,dpt_DOS_2015,Abdelrahim2017,Dolk-et-al-tac17,dpt_DOS_2015,Wang-Lemmon-aut11,Liu-Jiang-aut15}. Since  $A$ and $B$ in (\ref{system}) are unknown, we cannot apply these results and the challenge is thus to design the event-triggered controller directly based on some available data. In particular, we assume that we are given a set of data collected off-line 
\begin{equation} \label{dataset}
\mathbb D := \left\{ u(t),x(t),\dot x(t) \,:\, t\in\{0,T_s,\ldots,(T-1)T_s\}\right\},
\end{equation} 
where $T_s>0$ is a sampling time used 
for collecting the data and $T>0$ is the number of 
samples. Set $\mathbb D$ consists of input, state and state derivative data collected from the system with an experiment. This means  we have access to a set of input-state samples verifying $\dot x(t)=Ax(t)+Bu(t)+d(t)$ for $t\in\{0,T_s,\ldots,(T-1)T_s\}$. We consider data collected periodically only to ease the exposition: this is not required for the forthcoming results to hold, in other words aperiodic sampled data can very well be considered.

The problem of interest is to determine, using $\mathbb D$, a controller gain $K$ in \eqref{control}
and a triggering condition policy that defines the transmission instants  $\{t_k\}_{k \in \mathcal{N}}$ (not necessarily periodic) for each solution,  
such that the induced closed-loop system has suitable stability properties, as formalized in the sequel. We proceed by emulation for this purpose, in the sense that we will first design the stabilizing controller gain $K$ in the absence of sampling. Afterwards, sampled communications due to the network are taken into account and we will derive  triggering conditions to (approximately) preserve  stability. We first focus on the ideal case of noise-free data to introduce the required background and techniques in Section \ref{sec:exact}. These results are then extended  to the case where data are noisy in Section \ref{sec:noise}. We will explain how the followed approach can be used to develop various other triggering conditions in Section \ref{sect:other-triggering-rules}.


\begin{rem}[Data acquisition phase] The question of the selection $T_s$ and $T$ is  important for the forthcoming results and we will comment on this point later in the paper. Here, we remark that the sampling time $T_s$ used for collecting the data does not need to coincide 
with those generated by the triggering policy to be designed. \hfill $\Box$
\end{rem}
\begin{rem}[Measurement noise]\label{rem:measurement-noise}
All the results of this paper can  be extended to the case 
where the \emph{off-line} collected data are affected by measurement noise, i.e., to the case in which the process output 
(the measured signal) is $z=x+w$, with $w$ a noise signal. As shown in \cite[Section V-A]{de2019formulas}, this case reduces to input disturbances with suitable manipulations. When the \emph{on-line} measurements are corrupted by noise, the problem remains challenging even in the model-based setting, see, e.g., \cite{Abdelrahim2017,selivanov-fridman-tac15,Scheres-et-al-cdc2020,Scheres-et-al-automatica-submission}, and is left for future work. \hfill $\Box$
\end{rem}

\begin{rem}[State derivative measurements]\label{rem:derivative-data}
The \emph{off-line} computation of $\dot x$ in (\ref{dataset}) is error-prone.
We can explicitly account for errors in the computation of $\dot x$ by modelling these errors as a measurement noise. Error bounds when $\dot x$ is computed with Euler discretization are provided in \cite{berbCT2022}.
Alternatively, we can consider an \emph{integral} version of 
the relation $\dot x(t)=Ax(t)+Bu(t)+d(t)$ which permits to construct 
datasets that do not involve the computation of $\dot x$, see  Appendix \ref{app:int}.  \hfill $\Box$
\end{rem}

\section{Learning from noise-free data}\label{sec:exact}

\subsection{Assumption on set $\mathbb{D}$}


We consider throughout this section system (\ref{system}) with $d\equiv 0$ and a noise-free dataset  $\mathbb D$ in \eqref{dataset}. We define 
\begin{subequations}
\label{eq:data}
\begin{alignat}{4}
 U_0 & := \begin{bmatrix} u(0) & u(T_s) & \cdots & u((T-1)T_s)  \end{bmatrix} 
\in \mathbb R^{m \times T} \,, \label{eq:data1} \\
 X_0 & := \begin{bmatrix} x(0) & x(T_s) & \cdots & x((T-1)T_s)  \end{bmatrix} 
\in \mathbb R^{n \times T} \,, \label{eq:data2} \\
 X_1 & := \begin{bmatrix} \dot x(0) & \dot x(T_s) & \cdots & \dot x((T-1)T_s)  \end{bmatrix} 
\in \mathbb R^{n \times T}. \label{eq:data3} 
\end{alignat}
\end{subequations}
We assume the next condition holds on the \emph{richness} of the data in $\mathbb D$, which 
is discussed later on after Theorem \ref {thm:exact}. 

\begin{assumption} \label{ass:rich}
The matrix $\left[\begin{smallmatrix} U_0 \\ X_0 \end{smallmatrix}\right]$ has full row rank. \hfill  
$\Box$
\end{assumption}

Assumption \ref{ass:rich} can be easily checked for a given set $\mathbb D$.
For discrete-time systems, when $(u,d)$ is persistently exciting 
then Assumption \ref{ass:rich} holds \cite{willems2005}. For an extension of 
this result to continuous-time systems see \cite{Lopez2022}. 
We are ready to proceed with the design of the controller gain $K$ in (\ref{control}).


\subsection{Learning a feedback controller}

To design the feedback law, we follow the approach introduced in \cite{de2019formulas}, which gives simple formulas for controller design and returns a data-based
representation of the closed-loop dynamics that is useful to determine later on
the triggering policy. The next two results are taken from \cite{de2019formulas};
we recall them for convenience.  We start with an auxiliary result. 

\begin{lem} \label{lem:main}
Let Assumption \ref{ass:rich} hold and
consider any matrix $K \in \mathbb R^{m \times n}$. Then $A+BK=X_1G$ where
$G \in \mathbb R^{T \times n}$ 
is any solution to the system of equations 
\begin{equation} \label{eq:GK}
\begin{bmatrix} K \\ I \end{bmatrix} = \begin{bmatrix} U_0 \\ X_0 \end{bmatrix} G .
\end{equation}\hfill $\Box$
\end{lem} 
\smallskip

\emph{Proof.} 
By Assumption \ref{ass:rich}, there exists a $T \times n$ matrix $G$
such that \eqref{eq:GK} holds. Hence,
$A+BK=
[\begin{smallmatrix} B & A \end{smallmatrix}] [\begin{smallmatrix} K \\ I \end{smallmatrix}]=
[\begin{smallmatrix} B & A \end{smallmatrix}] [\begin{smallmatrix} U_0 \\ X_0 \end{smallmatrix}]G $,
where the second identity follows from \eqref{eq:GK}. The result
follows because the elements of $U_0,X_0$ and of $X_1$ satisfy the relation
$\dot x(t) = A x(t) + Bu(t)$, $t\in\{0,T_s\ldots,(T-1)T_s\}$, 
which, in compact form, gives $X_1 = AX_0 +BU_0$.
\hfill $\blacksquare$

We exploit Lemma \ref{lem:main} to design $K$ such that $A+BK$ is Hurwitz, despite the fact that $A$ and $B$ are unknown. As a result,  the origin of  system (\ref{system}) with $d\equiv 0$ in closed-loop with the continuous-time controller $u=Kx$ will be  globally exponentially stable, i.e., there exist $c_1\geq 1$ and $c_2>0$ such that any solution $x$ satisfies $|x(t)|\leq c_1 e^{-c_2 t}|x(0)|$ for any $t\geq 0$. We derive for this purpose a convex program, specifically a SDP.
As $A$ and $B$ are unknown, we use the fact that for any matrix $K\in\R^{n\times m}$, $X_1 G=A+BK$ where $X_1$ and $G$ are known from $\mathbb{D}$ in (\ref{dataset}) in view of Lemma \ref{lem:main},  to select suitable $K$ based on $U_0$, $X_0$ and $X_1$ in the next theorem. 

\begin{theorem} \label{thm:exact}
Consider stabilizable system \eqref{system} with $d \equiv 0$ and suppose Assumption \ref{ass:rich} holds. Consider the next SDP in the decision variable
$Y \in \mathbb R^{T \times n}$
\begin{subequations}
\label{eq:SDP}
\begin{alignat}{2}
X_1 Y + (X_1 Y)^\top \prec 0 , &  \label{eq:SDP1} \\
X_0 Y \succ 0 . & \label{eq:SDP2}
\end{alignat}
\end{subequations} 
SDP (\ref{eq:SDP}) is feasible and any solution $Y$ to \eqref{eq:SDP} is such that the
matrix $K=U_0 Y(X_0 Y)^{-1}$ renders 
$A+BK$ Hurwitz. As a result, the origin of system $\dot{x}(t)=(A+BK)x(t)$ is globally exponentially stable. \hfill $\Box$
\end{theorem} 

\emph{Proof.} We first show that \eqref{eq:SDP} is feasible.
Consider any matrix $K$ that makes $A+BK$ Hurwitz, which exists since
$(A,B)$ is stabilizable. 
By Lyapunov theory, there exists a symmetric matrix $S \succ 0$
such that $S (A+BK) + (A+BK)^\top S \prec 0$. Furthermore,
by Lemma \ref{lem:main} there exists a matrix $G$ that satisfies
$[\begin{smallmatrix} K \\ I \end{smallmatrix}]=
[\begin{smallmatrix} U_0 \\ X_0 \end{smallmatrix}]G$, which 
implies $A+BK=X_1G$. Hence, $Y:=GS^{-1}$ satisfies both \eqref{eq:SDP1}
and \eqref{eq:SDP2} as $X_0 G = I$ and $X_0 Y = S^{-1} \succ 0$. This shows that \eqref{eq:SDP} is feasible.

Concerning the second part of the statement, consider any solution $Y$ to \eqref{eq:SDP} and let $G:=YS$,
where we set $S:= (X_0 Y)^{-1}$. Note that $X_0 G=I$ and that 
$K=U_0 Y(X_0 Y)^{-1}$ can also be written as
$K=U_0G$, as \eqref{eq:GK} holds.
By combining \eqref{eq:SDP1} and \eqref{eq:SDP2}, and because  
$G=YS$ and $A+BK=X_1G$, we conclude that $S (A+BK)  + (A+BK)^\top S \prec 0$.
Thus $A+BK$ is Hurwitz as $S$ is symmetric and $S\succ0$. The global exponential stability of the origin of $\dot{x}(t)=(A+BK)x(t)$ then follows, see \cite[Theorem 8.2]{hespanha2018linear}. \hfill $\blacksquare$

The decision variable in Theorem \ref{thm:exact} is $Y$ (and not $G$) and $Y(X_0 Y)^{-1}=G$. This change of variable is instrumental to arrive at a convex formulation of the design program as shown in the proof of Theorem \ref{thm:exact}. We also notice from the proof of Theorem \ref{thm:exact} that 
Assumption \ref{ass:rich} is not needed for the second part of the result, which implies that there might exist a solution to \eqref{eq:SDP} even though  Assumption \ref{ass:rich}
does not hold. This fact has been pointed out in the discrete-time case in
\cite{van2020data}. Nevertheless, having $\left[\begin{smallmatrix} U_0 \\ X_0 \end{smallmatrix}\right]$ full row rank 
gives certain advantages as \emph{any}
stabilizing controller can be parametrized through the data. 
This is useful when we search for a controller that satisfies extra desirable properties. 

Theorem \ref{thm:exact}  provides a way to design the state feedback controller gain $K$ in absence of network (in the noise-free case). We now  move to the next step of the emulation approach that is to take sampling into account and to design the triggering condition.

\subsection{Learning a triggering policy}\label{subsect:trig-rule-noisefree} 


We write the  closed-loop system under the control law \eqref{control} as
\begin{equation} \label{eq:closed-loop}
\dot x(t) = (A+BK) x(t) + BK e(t)
\end{equation}
where  
\begin{equation} \label{eq:e}
e(t) := x(t_k) - x(t), \quad t \in [t_k,t_{k+1}),
\end{equation}
represents the sampling-induced error, that is the mismatch between the last value of the state transmitted to the controller and its current value. As customary in the event-triggered control literature, the idea is to regard $e$
as a disturbance to the nominal dynamics $\dot x(t) = (A+BK) x(t)$, which is then controlled via the triggering condition so that stability is preserved despite sampling.

We develop a data-based version of the approach proposed in \cite{Tabuada07} for this purpose; the extension to other triggering conditions is addressed in Section \ref{sect:other-triggering-rules}. 
We introduce for this purpose the next parameterized matrix and the vector $z$
\begin{equation} \label{eq:Psi}
\Psi(\sigma) := \begin{bmatrix} -\sigma^2 I & \mathbf{0} \\ \mathbf{0} & I \end{bmatrix}, \quad 
z := \begin{bmatrix} x \\ e \end{bmatrix},
\end{equation}
with $\sigma > 0$ a design parameter to be determined.
The sampling times are defined as follows: $t_0:=0$ and
\begin{equation} \label{eq:triggering}
t_{k+1} = \left\{ 
\begin{array}{l}
\inf \{t \in \mathbb R : t > t_k \text{ and } z(t)^\top \Psi(\sigma) z(t) =0  \}  \\[0.1cm]
\hfill \text{if } x(t_k)\neq0,  \\[0.2cm]
 + \infty \hfill \text{otherwise.}
\end{array}
\right.
\end{equation}
This logic ensures by design that $z^\top \Psi(\sigma) z \leq 0$  
along the solutions to \eqref{eq:closed-loop}, \eqref{eq:e}, \eqref{eq:triggering} as long as they
exist; we shall prove later on that the sequence of 
sampling instants does not result in an accumulation point, which guarantees that a solution to \eqref{eq:closed-loop}, \eqref{eq:e}, \eqref{eq:triggering}  exists for all times (and is unique).
This logic may provide asynchronous and sporadic control updates. In fact, it owes its popularity thanks to this latter feature besides
its conceptual simplicity. 

Closed-loop stability depends 
on $\sigma$, which must be chosen sufficiently small to  control the norm of error $e$. How small $\sigma$ should be is system-dependent in the sense that the matrices $A$ and $B$
determine which values of $\sigma$ are allowed, see \cite[Section IV]{Tabuada07}.
Hereafter, we propose a data-based method to determine $\sigma$. 

Consider any controller gain $K$ computed via Theorem \ref{thm:exact} and let 
$V(x)=x^\top S x$ for any $x\in\R^{n}$ with $S=(X_0Y)^{-1}$, which is positive definite by Theorem \ref{thm:exact}. Function $V$ serves as a Lyapunov 
function for the closed-loop system ignoring sampling. In particular we have, for any $z=(x,e)\in\R^{2n}$, 
\begin{equation} \label{eq:V_exact}
\left\langle \nabla V(x),(A+BK)x+BKe\right\rangle = z^\top 
\underbrace{\begin{bmatrix}
-Q & SX_1L \\ (SX_1L)^\top & \mathbf{0}
\end{bmatrix}}_{\displaystyle =: M} z,
\end{equation} 
where 
\begin{equation} \label{eq:Q}
Q := - (SX_1G)^\top - S X_1G,
\end{equation} 
which is positive definite in view of Theorem \ref{thm:exact} and its proof (recall that $X_1 G=A+BK$ by Lemma \ref{lem:main}). Matrix  $L$ in (\ref{eq:V_exact}) is any solution to the system of equations
\begin{equation} \label{eq:L}
\begin{bmatrix} K \\ \mathbf{0} \end{bmatrix} = \begin{bmatrix} U_0 \\ X_0 \end{bmatrix} L ,
\end{equation}
which exists under Assumption \ref{ass:rich}. Note that \eqref{eq:L}  
implies that $BK=X_1L$.
Recall now that the event-triggering rule \eqref{eq:triggering} ensures that
$z^\top \Psi(\sigma) z\leq 0$ along the solutions to \eqref{eq:closed-loop}, \eqref{eq:e}, \eqref{eq:triggering}. Hence, the global exponential stability of the origin of closed-loop system \eqref{eq:closed-loop}, \eqref{eq:e} is ensured provided we select $\sigma$ such that the  
following implication\footnote{Relations involving quadratic inequalities are well-known in 
optimization and control theory, usually with the term ``$S$-procedure". Recently, there has been interest 
for the $S$-procedure also in the context of data-driven control, see \cite{henk-ddctr-uncer} 
for a recent discussion and new results.} is satisfied for some $\varepsilon>0$ and any $z\in\R^{2n}$  

\begin{equation} \label{eq:SDP_triggering_0}
\begin{array}{rlll}
z^\top \Psi(\sigma) z \leq 0 
& \Longrightarrow &   z^\top M z \leq -\varepsilon|z|^2.
 \end{array}
\end{equation} 
This implication ensures that $V$ exponentially decreases along any solution to \eqref{eq:closed-loop}, \eqref{eq:e}, \eqref{eq:triggering} in view of (\ref{eq:V_exact}) and the facts that $|x|\leq |z|$ for any  $z=(x,e)\in\R^{2n}$ and that $V$ is a positive definite quadratic function. This implies that $x=0$  is  globally exponentially stable for system \eqref{eq:closed-loop}, \eqref{eq:e}, \eqref{eq:triggering} (again, provided Zeno phenomenon does not occur). The next theorem provides a data-based condition to select $\sigma$ in (\ref{eq:triggering}) to ensure the desired stability property.




\begin{theorem} \label{thm:exact_sampling}
Suppose Assumption \ref{ass:rich} holds  and consider system (\ref{eq:closed-loop}), (\ref{eq:e}), (\ref{eq:triggering}) with $K=U_0Y(X_0Y)^{-1}$, $Y$ being any solution to \eqref{eq:SDP}. Let  $\mu,\,\sigma>0$ be such that the next SDP (in the decision variables 
$\mu$ and 
$\sigma^2$) is satisfied 
\begin{equation}
\label{eq:SDP_triggering}
\mu M - \Psi(\sigma) \prec 0,
\end{equation}
with $S=(X_0Y)^{-1}$ and $L$  as in \eqref{eq:L}. 
Then the following holds.
\begin{itemize}
\item[(a)] There exists a \emph{global minimum inter-event time}, in particular, for any solution $x$ to the system, the sequence of transmission times $\{t_k\}_{k\in\mathcal{N}}$ satisfies $t_{k+1}-t_k \geq \tau(\sigma)$ for every $k \in \mathcal{N}=\mathbb N_0$ where $\tau(\sigma):=\frac{1}{\alpha} \frac{\sigma}{1+\sigma}$ with $\alpha:= \max\{\|X_1G\|,\|X_1L\|\}$. 
\item[(b)] The origin of the system is  globally exponentially stable.
\hfill $\Box$
\end{itemize}  
\end{theorem}  


\emph{Proof.} We first show the feasibility of \eqref{eq:SDP_triggering}. Let $\sigma:=\sqrt{\mu/c}$
with $c>0$ any sufficiently large constant such that $-Q + I/c \prec 0$ 
where $Q$ comes from (\ref{eq:Q}); clearly, this $c$ exists because $Q \succ 0$.
For such a choice of $\sigma$, we have $-\mu Q + \sigma^2 I = \mu (-Q+I/c) \prec 0$ 
for every $\mu>0$. Therefore, for this choice of $\sigma$, \eqref{eq:SDP_triggering}
is equivalent by a Schur complement to the two conditions
\begin{equation} 
\begin{array}{rl}
\mu (-Q+I/c) \prec 0 \text{ \,\,and} \\[0.1cm]
-I + \mu (SX_1L)^\top (Q - I/c)^{-1} (SX_1L) \prec 0
 \end{array}
\end{equation} 
which are jointly satisfied for $\mu$ sufficiently small. 

We now prove item (a) of Theorem \ref{thm:exact_sampling}. We follow the same steps as \cite[Section III]{Tabuada07} adapted to the data-based context of this work. Let $x$ be a solution to (\ref{eq:closed-loop}), (\ref{eq:e}), (\ref{eq:triggering}).
The claim is trivial when $x(0)=0$. When $x(0)\neq0$, $x(t)\neq 0$ for any $t$ in the domain of the solution, see \cite[Proposition 4]{Postoyan_mystery}. Moreover, we also have from \cite[Proposition 1]{Postoyan_mystery} that the number of jumps is infinite, i.e., $\mathcal{N}=\mathbb{N}_0$.  
Let $k\in\mathbb{N}_0$, we have for any $t\in [t_k,t_{k+1})$, 
\begin{subequations}
\label{eq:Tab_MIET}
\begin{alignat}{2}
\frac{d}{dt} \frac{|e(t)|}{|x(t)|} = & \,
\frac{e(t)^\top \dot e(t)}{|e(t)| |x(t)|} - \frac{x(t)^\top \dot x(t) |e(t)|}{|x(t)|^3} \\
\leq & \, \left( 1+ \frac{|e(t)|}{|x(t)|} \right) \frac{|\dot x(t)|}{|x(t)|} \\
\leq & \, \alpha \left( 1+ \frac{|e(t)|}{|x(t)|} \right)^2 
\end{alignat}
\end{subequations} 
with $\alpha := \max\{\|X_1G\|,\|X_1L\|\}$,
where the last inequality is obtained by recalling that
$A+BK=X_1G$ and $BK=X_1L$, so that
$\dot x = X_1G x + X_1Le$ which implies $|\dot x| \leq \alpha (|x|+|e|)$.
Therefore, the time needed for $|e|/|x|$ to reach $\sigma$ is lower bounded by $\tau(\sigma)$,  
the time needed for $\phi$ to grow from $0$ (the value of $|e|/|x|$ right after each jump) to $\sigma$, where $\phi$ is the solution to
the differential equation $\dot \phi = \alpha (1+\phi)^2$.
The expression of $\tau(\sigma)$ is given by 
\begin{equation} \label{eq:tau}  
\tau(\sigma) := \frac{1}{\alpha}  \frac{\sigma}{1+\sigma}.
\end{equation} 
Thus, $t_{k+1}-t_k \geq \tau(\sigma)$. Since the solution $x$ and the number of jump $k$ have been arbitrarily selected, we have proved that  $\tau(\sigma)$ is a global minimum inter-event time. This property implies that any solution to  (\ref{eq:closed-loop}), (\ref{eq:e}), (\ref{eq:triggering}) is complete, and the linearity of the flow vector field  and (\ref{eq:triggering}) ensure the uniqueness of the solution for each initial condition, see \cite{cortesDDS}.

Finally, as \eqref{eq:SDP_triggering} holds, there exists $\varepsilon>0$ such that $M\prec 1/\mu\,\Psi(\sigma)-\varepsilon I$. 
Hence, the satisfaction of \eqref{eq:SDP_triggering} implies that \eqref{eq:SDP_triggering_0} holds. Equation \eqref{eq:triggering} enforces 
$z^\top \Psi(\sigma) z \leq 0$ along the solutions to (\ref{eq:closed-loop}), (\ref{eq:e}), (\ref{eq:triggering}). Hence, for any solution $x$ to (\ref{eq:closed-loop}), (\ref{eq:e}), (\ref{eq:triggering}), 
$\dot V(x(t))=z(t)^\top M z(t) \leq -\varepsilon|z(t)|^2$ for any $t\geq 0$; note that $V(x)$ is not affected by jumps at the triggering instants. Since $V$ is quadratic and  positive definite, we conclude that item (b) of Theorem \ref{thm:exact_sampling} holds by resorting to standard Lyapunov arguments. 
\hfill $\blacksquare$

The condition in (\ref{eq:SDP_triggering}), which is linear in the  decision variables $\mu > 0$ and $\sigma^2>0$, provides a data-based condition to design $\sigma$ so that the induced event-triggered controlled system enjoys a  global exponential stability property as well as the existence of  global minimum inter-event time $\tau(\sigma)$. We emphasize that the expression of $\tau(\sigma)$ in item (a) of Theorem \ref{thm:exact_sampling} only depends on  experimental data. As a result,  different experiments and even different solutions to
the SDP in \eqref{eq:SDP_triggering} may lead to a different value for $\tau(\sigma)$. This expression of $\tau(\sigma)$ can, by the way, be used as a  sampling period for \emph{periodic} implementations. 

\begin{rem}\label{rem:decay-on-V}
Condition (\ref{eq:SDP_triggering}) may be used to maximize $\sigma$, which may help enlarging the inter-event times thereby  reducing the number of transmissions, see  \cite[Proposition 3]{Postoyan_mystery} for further insights on the relationship between the inter-event times and $\sigma$. In particular, we can implement the next SDP in place of \eqref{eq:SDP_triggering} in this case 
\begin{subequations}
\label{eq:SDP_triggering_max}
\begin{alignat}{2}
\textrm{maximize}_{\mu,\sigma^2} \,\,
& \sigma^2  \\ 
\textrm{subject to} \,\,
& \eqref{eq:SDP_triggering}.
\end{alignat}
\end{subequations} 
In this regard, it is interesting to note that we may achieve
parsimonious sampling while controlling the performance. In fact,
suppose that we replace \eqref{eq:SDP_triggering} with  
$\mu M_\gamma - \Psi(\sigma) \prec 0$, where
$M_\gamma$ is equal to $M$ except for $Q$, which is replaced by $(1-\gamma)Q$
where $\gamma \in (0,1)$ is a design parameter. 
Theorem \ref{thm:exact_sampling} remains valid 
with the difference that instead of $\left\langle \nabla V(x),(A+BK)x+BKe\right\rangle  \leq - \varepsilon|z|^2\leq - \varepsilon|x|^2$ for some  $\varepsilon>0$, we will now have $\left\langle \nabla V(x),(A+BK)x+BKe\right\rangle \leq -\gamma x^\top Q x$ for any $z\in\R^{2n}$. 
In this way, we can maximize $\sigma$ while keeping  
a fixed \emph{guaranteed} level of performance in terms of Lyapunov decay along the solutions.\hfill $\Box$
\end{rem}

\begin{figure*}[!t]
\caption{Results for Example \ref{ex:1}. Left: State trajectories.
Middle: Behavior of $|e(t)|$ and $\sigma |x(t)|$;  
a new sampling is triggered when $|e(t)|=\sigma |x(t)|$. 
Right: Behavior of the inter-sampling times $t_{k+1}-t_k$. 
The minimum inter-sampling time observed in simulation is $0.1371$, which is not much 
larger than the theoretical lower bound $\tau(\sigma)=0.1184$.
Interestingly, the inter-sampling times exhibit an oscillatory behavior consistently with \cite[Section IV.A]{Postoyan_mystery}.} \label{fig:ex1}
{\includegraphics[width=5cm]{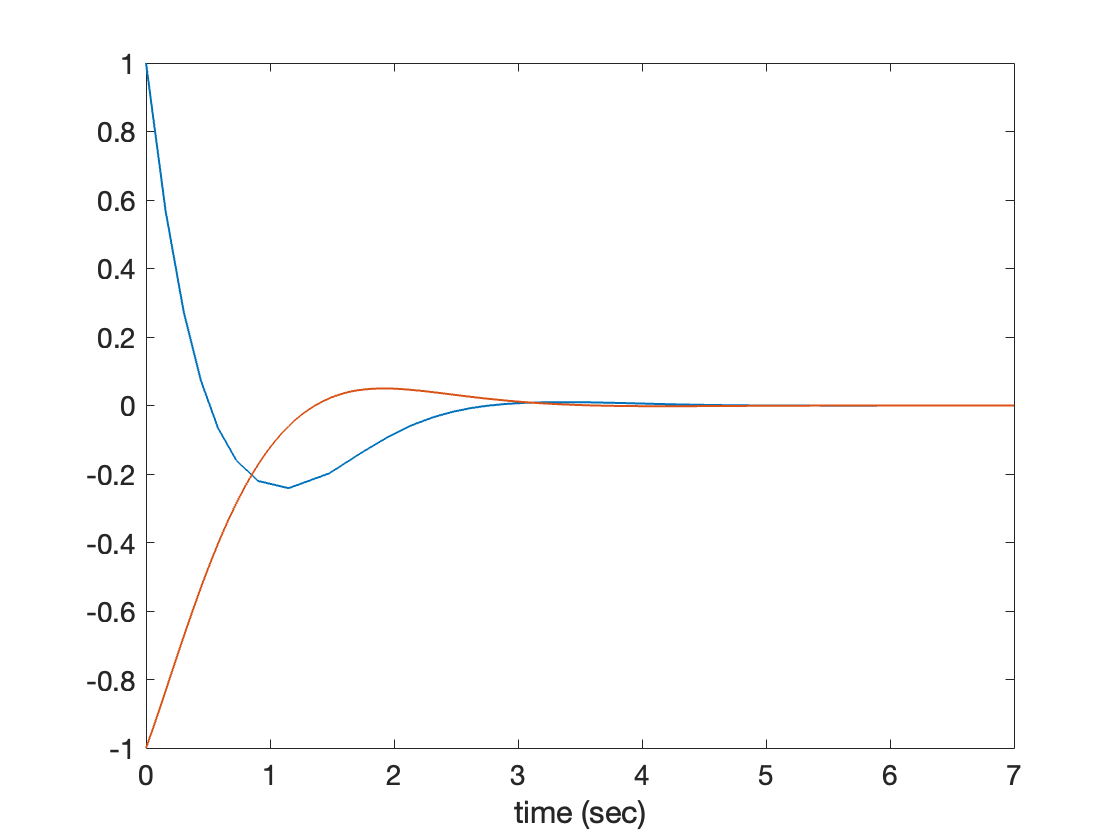}} \hspace*{-0.7cm}
{\includegraphics[width=5cm]{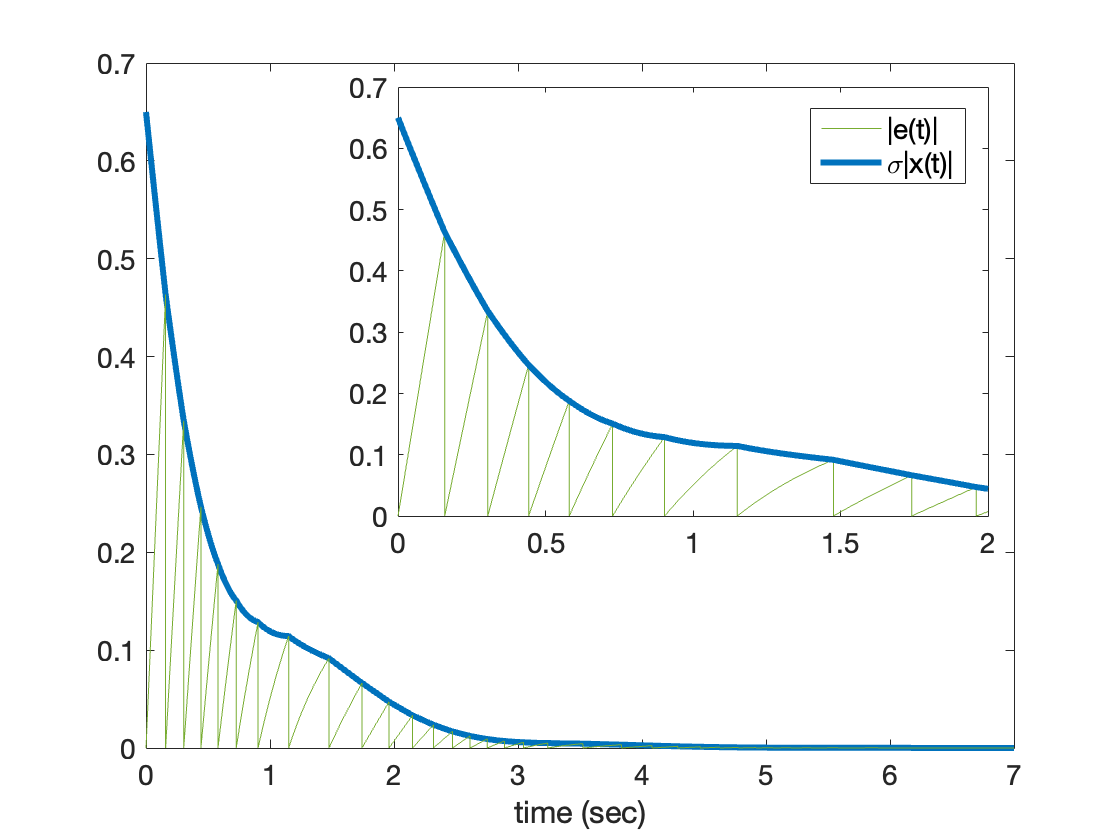}} \hspace*{-0.7cm} 
{\includegraphics[width=5cm]{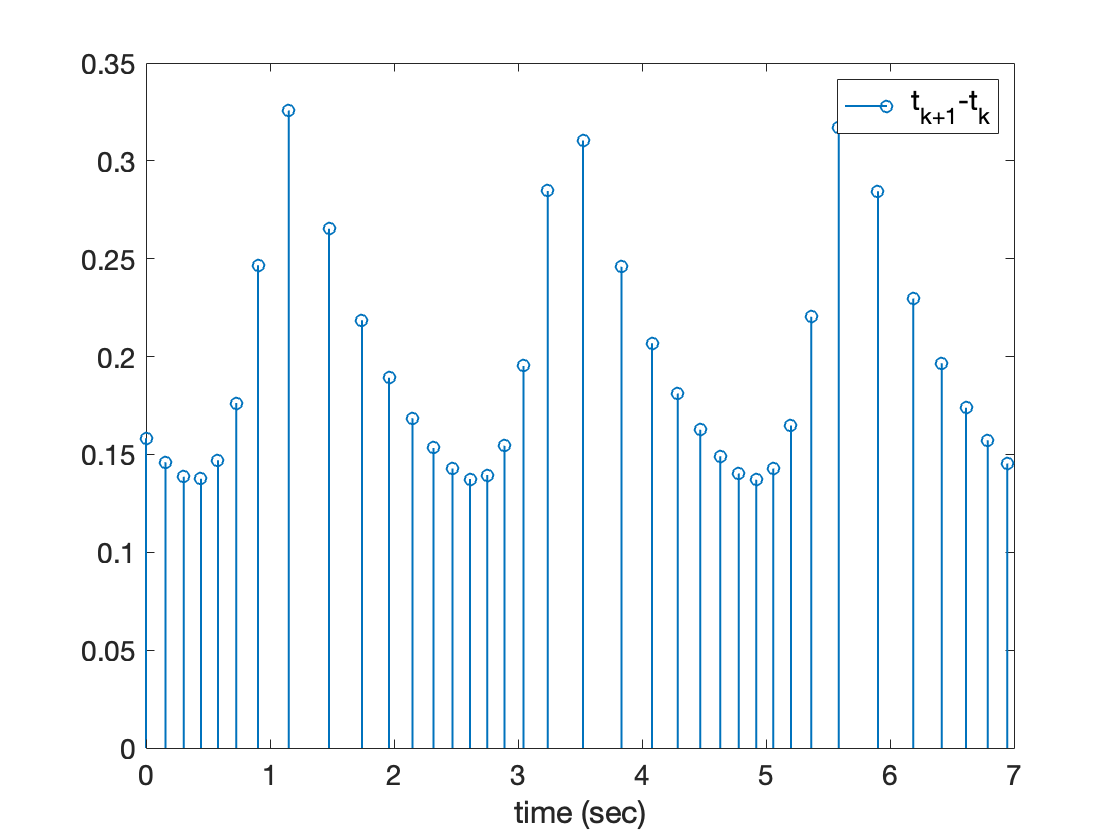}} 
\setcounter{equation}{\value{equation}}
\end{figure*}

\begin{example} \label{ex:1}
Consider  system \eqref{system} with $d \equiv 0$ and
\begin{equation*}
A = \begin{bmatrix} 0 & 0 \\  -1 & -2 \end{bmatrix}, \quad
B = \begin{bmatrix} 1 \\ 0 \end{bmatrix} ,
\end{equation*}
which is marginally stable. We collect the data by running an experiment with input uniformly 
distributed in $[-1,1]$ and initial state within the same interval.  
The data are collected by keeping the input constant, 
in a sample-and-hold fashion, during each interval $[iT_s,(i+1)T_s)$ 
with $i=0,1,\ldots,T-1$ and $T_s=0.1$.
We collect $T=10$ samples which is sufficient for Assumption \ref{ass:rich} to hold.
Note that theoretically we need $T \geq n+m$, so that even $T=3$ samples may be sufficient.
We solve SDP \eqref{eq:SDP} and we obtain $K = \begin{bmatrix} -0.9839  & 1.7489 \end{bmatrix}$.
We then solve \eqref{eq:SDP_triggering_max}, which gives $\sigma=0.4595$. According to 
\eqref{eq:tau},  
the minimum inter-event time $\tau(\sigma)$ is equal to $0.1184$, which is only an estimate of the actual minimum inter-event times. Figure \ref{fig:ex1} reports simulations results. 

We have not observed critical issues regarding the choice of the sampling time $T_s$ which is
used for collecting the data, except when it becomes extremely small ($T_s < 0.001$).
This is not surprising: in this case the samples $x(iT_s)$ and $u(iT_s)$ are
approximately constant.
In particular, \eqref{eq:SDP} becomes infeasible as soon as $X_0$ (approximately) looses rank.
\hfill $\Box$
\end{example}

\section{Learning from noisy data}\label{sec:noise}

In this section, we consider system \eqref{system} with $d$ not identically equal to $0$. With noisy data, the analysis and the design of the event-triggered controller become sensibly more involved, and the reason
is twofold. First, noisy data prevent us from obtaining an exact data-based 
representation of the closed-loop behavior as in Lemma \ref{lem:main}. In fact, more generally,
noisy data prevent \emph{exact} identification, whether this involves the closed-loop dynamics
(Lemma \ref{lem:main}) or open-loop dynamics (the matrices $A$ and $B$).
Second, disturbances impact not only the learning phase but, subsequently, 
also the behavior of the closed-loop system. In particular, Zeno phenomenon may arise in this case with the triggering rule built in Section \ref{subsect:trig-rule-noisefree} even for arbitrarily small disturbances, as demonstrated in \cite{Borgers-Heemels-tac14}.

Because we do not assume that the disturbance $d$ is vanishing, we can no longer hope to prove  that $x=0$ is  globally exponentially stable for the considered closed-loop system as in Theorems \ref{thm:exact} and \ref{thm:exact_sampling}. We will establish instead a (global) practical exponential ISS property, i.e., that there exist $c_1\geq 1$, $c_2,c_3,c_4>0$ such any solution $x$ to  (\ref{system}) with input disturbance $d$ in closed-loop with (\ref{control}) satisfies
\begin{equation}
    \begin{array}{rlll}
    |x(t)|  \leq  c_1 e^{-c_2 t}|x(0)|+c_3 \|d\|_{[0,t]}+c_4\nu,
    \end{array}\label{eq:exp-iss-practical}
\end{equation}
where $\nu\geq 0$ is an extra tuning parameter we will introduce to design the triggering rule. When (\ref{eq:exp-iss-practical}) holds with $\nu=0$, we say that the system is exponentially ISS for system \eqref{system}, \eqref{control}.

As in Section \ref{sec:exact}, we first design the control law ignoring sampling and we then present the design of the triggering policy. Two triggering techniques are presented and discussed in this section, whether these lead to (practical) exponential ISS. 
Again, the extension of the presented results to other triggering conditions is addressed in Section \ref{sect:other-triggering-rules}.

\subsection{Learning a feedback controller}

The first step
is to derive an analogue of Lemma \ref{lem:main}.
Suppose we perform an experiment on the 
system, and we collect state and input samples\footnote{See, again, Remarks \ref{rem:measurement-noise} and \ref{rem:derivative-data}   for explanations regarding the possible presence of noise affecting these data.} that satisfy $\dot x(t)=Ax(t)+Bu(t)+d(t)$,
$t\in\{0,T_s,\ldots,(T-1)T_s\}$. 
The samples are then grouped into the data matrices $U_0, X_0, X_1$ 
as in \eqref{eq:data}. Let 
\begin{equation} \label{eq_D0}
D_0 := \begin{bmatrix} d(0) & d(T_s) & \cdots & d((T-1)T_s)  \end{bmatrix}  
\end{equation}
be the \emph{unknown} data matrix made of the off-line samples of $d$. We have the next result.

\begin{lem} \label{lem:main2}
Let Assumption \ref{ass:rich} hold and
consider any matrix $K \in \mathbb R^{m \times n}$. Then $A+BK=(X_1-D_0)G$ where
$G \in \mathbb R^{T \times n}$ 
is any solution to the system of equations \eqref{eq:GK}. 
\end{lem} 

\emph{Proof.} 
By Assumption \ref{ass:rich}, there exists a $T \times n$ matrix $G$
such that \eqref{eq:GK} holds. Hence
$A+BK=
[\begin{smallmatrix} B & A \end{smallmatrix}] [\begin{smallmatrix} K \\ I \end{smallmatrix}]=
[\begin{smallmatrix} B & A \end{smallmatrix}] [\begin{smallmatrix} U_0 \\ 
X_0 \end{smallmatrix}]G $,
where the second identity follows from \eqref{eq:GK}. The result
follows because the elements of $U_0,X_0,X_1$ and $D_0$ satisfy the relation
$\dot x(t) = A x(t) + Bu(t)+d(t)$, $t\in\{0,T_s,\ldots,(T-1)T_s\}$
which, in compact form, gives $X_1 = AX_0 +BU_0+D_0$.
\hfill $\blacksquare$

By Lemma \ref{lem:main2}, the closed-loop dynamics 
can be written as $\dot x(t)=(A+BK)x(t)=(X_1-D_0)G_1x(t)$, which depends on the unknown matrix $D_0$. As a result, \eqref{eq:SDP} 
no longer provides stability guarantees. In fact, the constraint \eqref{eq:SDP1}
ensures that $X_1G_1$ is Hurwitz. However,
the matrix of interest is now $\Phi:=(X_1-D_0)G_1$ in view of Lemma \ref{lem:main2}, and  $X_1G_1$ Hurwitz does not imply that $\Phi$ is Hurwitz. To be able to prove an ISS property for system (\ref{system}) in closed-loop with $u=Kx$ (ignoring sampling),
we need to modify \eqref{eq:SDP1} accounting for the uncertainty induced by $D_0$. 
An effective way to achieve this is to ensure 
that $(X_1-D)G_1$ satisfies a Lyapunov equation with a common (symmetric) Lyapunov matrix $Y\succ 0$  \emph{for all}
the matrices $D$ in a given set $\mathcal D$ to which 
$D_0$ is deemed to belong as formalized next; this approach can in fact be viewed as a robust control approach. We thus introduce the set
\begin{equation} \label{eq:noise_model}
\mathcal D := \{ D \in \mathbb R^{n \times T}: 
D D^\top \preceq \Delta \Delta^\top \},
\end{equation}  
where $\Delta \in \mathbb R^{n \times s}$ is a known matrix, and we make the next assumption.
\begin{assumption} \label{ass:D0} $D_0 \in \mathcal D$. \hfill $\Box$
\end{assumption}

The choice of set $\mathcal D$ reflects our prior information or guess about disturbance $d$.
For instance, if we know that there exists $\delta>0$ such that $\|d\|_{\infty} \leq \delta$ like in \cite{qi-et-al-tie2022},
then we can take $\Delta :=\delta \sqrt{T} I$, where we recall that $T$ is the number of samples. Stochastic disturbances can also be accounted for 
(possibly with other choices of $\Delta$) as advocated in \cite{de2021low,dpte-nonl-canc}.

We aim at enforcing the next inequality instead of \eqref{eq:SDP1}
\begin{equation} \label{eq:2SDP_noisy} 
(X_1-D) Y + Y^\top (X_1-D)^\top + \Omega 
\prec 0  \quad \forall D \in \mathcal D
\end{equation}
where $\Omega\succ0$ is a free design parameter. 
By ensuring \eqref{eq:2SDP_noisy}, we will see that system $ \dot x(t)=(X_1-D)Gx(t)+d(t)$ is exponentially ISS given any $D \in \mathcal D$, and therefore for $D=D_0$ in view of Assumption \ref{ass:D0}. Large sets $\mathcal D$ make the condition $D_0 \in \mathcal D$ easier to hold
but make \eqref{eq:2SDP_noisy} more difficult to satisfy. 


Condition \eqref{eq:2SDP_noisy} cannot be implemented directly as it involves 
infinitely many constraints. 
The next result provides a tractable (convex) sufficient condition for \eqref{eq:2SDP_noisy} to hold,
thereby extending Theorem \ref{thm:exact} to the case of noisy data.

\begin{theorem} \label{thm:approx}
Consider stabilizable system  \eqref{system} and suppose the following holds.
\begin{enumerate}
\item[(i)] Assumptions \ref{ass:rich} and \ref{ass:D0} hold 
with $\Delta$ given. 
\item[(ii)] Let $\Omega \succ 0$,  there exist $\epsilon > 0$ and
$Y \in \mathbb R^{T \times n}$ such that 
\begin{subequations}
\label{eq:2SDP}
\begin{alignat}{2}
\left[ \begin{array}{cc}
X_1Y + (X_1Y)^\top + \Omega + \epsilon \Delta \Delta^\top & Y^\top \\[0.1cm] 
Y & -\epsilon I
\end{array}
\right] \prec 0, \label{eq:2SDP1} & \\
X_0 Y \succ 0 . & \label{eq:2SDP2}
\end{alignat}
\end{subequations} 
\end{enumerate}
Let $K=U_0 Y(X_0 Y)^{-1}$, then $\dot{x}(t)=(A+BK)x(t)+d(t)$ is exponentially ISS. \hfill $\Box$
\end{theorem} 

The proof of Theorem \ref{thm:approx} rests on the next result,
which we prove in Appendix \ref{sec:petersen} to not overload the exposition. 

\begin{lem} \label{lem:Petersen}
Suppose there exist a scalar $\epsilon > 0$ and a
matrix $Y \in \mathbb R^{T \times n}$ such that \eqref{eq:2SDP1} holds.
Then \eqref{eq:2SDP_noisy} holds. \hfill $\Box$
\end{lem} 


\emph{Proof of Theorem \ref{thm:approx}.} 
Let $G:=YS$ where $S:= (X_0 Y)^{-1}$. Note that $X_0 G=I$ and that 
$K=U_0 Y(X_0 Y)^{-1}$ can also be written as
$K=U_0G$. Accordingly, the identity \eqref{eq:GK} holds and
Lemma \ref{lem:main2} implies that $A+BK=(X_1-D_0)G$.
We now prove  that $(X_1-D_0)G$ is Hurwitz. By Lemma \ref{lem:Petersen},
\eqref{eq:2SDP_noisy} holds and this implies that
$S (X_1-D)G  + G^\top (X_1-D)^\top S \prec 0$
for all $D \in \mathcal D$. Thus, $(X_1-D)G$ is
Hurwitz for all $D \in \mathcal D$. As $D_0\in\mathcal{D}$ by Assumption \ref{ass:D0}, we 
thus have $A+BK$ Hurwitz, which implies that $\dot{x}(t)=(A+BK)x(t)+d(t)$ is exponentially ISS, see \cite[Example  10.4.1]{Isidori:nonlinear-control-systems-2}\hfill $\blacksquare$

\subsection{Learning a triggering policy}\label{subsect:triggering-rule-noisy}
\subsubsection{Mixed triggering condition}\label{subsubsect:mixed-triggering-rule}
We now take into account sampling. The dynamics of the closed-loop system under the control law \eqref{control} becomes
\begin{equation} \label{eq:closed-loop2}
\dot x(t) = (A+BK) x(t) + BK e(t) + d(t),
\end{equation}
where we recall that 
\begin{equation} \label{eq:e-noisy}
e(t) := x(t_k) - x(t), \quad t \in [t_k,t_{k+1}).
\end{equation}
There are two main obstacles compared to the noise-free case.
First, as we will see shortly, like $A+BK$ depends on $D_0$ in the data-based representation according to Lemma \ref{lem:main2},  so do $BK$ and $M$ in \eqref{eq:V_exact}.
This means that we cannot implement the inequality \eqref{eq:SDP_triggering} 
used in Section \ref{subsect:trig-rule-noisefree} to design the triggering policy parameter $\sigma$. Second, we need to modify 
\eqref{eq:triggering} in order to ensure the existence of a global minimum inter-event time. 

We start with the last point, and, inspired by the model-based results in \cite{Borgers-Heemels-tac14}, we modify \eqref{eq:triggering}
as: $t_0=0$ and  
\begin{equation} \label{eq:triggering2}
t_{k+1} = \inf \{t \in \mathbb R : t > t_k \text{ and } |e(t)| = \sigma|x(t)| + \nu  \},
\end{equation}
where $\sigma > 0$ is a design parameter to be determined, while $\nu>0$ is 
an arbitrary constant. 
Triggering rule (\ref{eq:triggering2}) is commonly called mixed (relative-absolute) threshold policy in the event-triggered control literature \cite{Borgers-Heemels-tac14}. 

Next we introduce the following quantities in place of \eqref{eq:Psi} and  matrix $M$ in \eqref{eq:V_exact}, respectively,
\begin{equation} \label{eq:Psi2}
\overline \Psi(\sigma) := \begin{bmatrix} -2 \sigma^2 I & \mathbf{0} \\ \mathbf{0} & I \end{bmatrix}, 
\end{equation}
and
\begin{equation} \label{eq:MD0}
\overline M(D) := \begin{bmatrix} 
-\displaystyle \frac{S \Omega S}{2}  & S(X_1-D)L \\[0.3cm] 
(S(X_1-D)L)^\top & \mathbf{0} \end{bmatrix}, 
\end{equation}
where $S:=(X_0Y)^{-1}$ and $Y$ results from 
Theorem \ref{thm:approx} and where $L$ is any solution to 
\eqref{eq:L}, which exists under Assumption \ref{ass:rich}. 

We are going to show is that, instead of \eqref{eq:SDP_triggering},
we can consider the condition $\mu \overline M(D_0) - \overline \Psi(\sigma) \preceq 0$,
which can be conveniently cast as a SDP. Consider any solution to \eqref{eq:2SDP}
and let $V(x):=x^\top S x$ for any $x\in\R^{n}$ with $S=(X_0Y)^{-1}$. Like in the noise-free case, 
$V$ is an ISS  Lyapunov function for the nominal dynamics, in particular we now have, for any $z=(x,e)\in\R^{2n}$ and $d\in\R^{n}$,
\begin{subequations}
\label{eq:V_approx}
\begin{alignat}{4} 
& \left\langle \nabla V(x),(A+BK)x+BKe+ d\right\rangle \nonumber \\ 
& =\, 2 \left[ (A+BK) x + BK e+d \right]^\top S x \\
& =\, 2 \left[ (X_1-D_0)Gx+ (X_1-D_0)L e+d \right]^\top S x \\
& \leq\, -x^\top S \Omega S x + 2 \left[  (X_1-D_0)L e + d \right]^\top S x \\
& =\, - x^\top \left( \frac{S \Omega S}{2} \right) x + z^\top \overline M(D_0) z + 2 d^\top S x.\label{eq:V_approx-last}
\end{alignat} 
\end{subequations} 
The second equality follows because $A+BK=(X_1-D_0)G$
and $BK=(X_1-D_0)L$ (see \eqref{eq:GK} and \eqref{eq:L}),
while the inequality is a consequence of \eqref{eq:2SDP_noisy}.
Finally, the last equality follows from the definition of $\overline M(D)$.
Since $S \Omega S= (S \Omega S)^\top\succ 0$, if we ensure that 
$z^\top \overline M(D_0) z \leq c$ along the solutions to (\ref{eq:closed-loop2})-(\ref{eq:triggering2}) for some positive constant
$c$ (and Zeno phenomenon does not occur), then we can conclude that (\ref{eq:exp-iss-practical}) holds for system (\ref{eq:closed-loop2})-(\ref{eq:triggering2}). This is formalized in the next theorem. 

\begin{theorem} \label{thm:practical}
Suppose the following holds.
\begin{enumerate}
\item[(i)] Assumptions \ref{ass:rich} and \ref{ass:D0} are verified with $\Delta$ given.
\item[(ii)] Let $\Omega \succ 0$, \eqref{eq:2SDP} holds and $K=U_0 Y (X_0 Y)^{-1}$
is the resulting controller gain as in Theorem \ref{thm:approx}.
\item[(iii)] There exist $\mu, \epsilon, \sigma > 0$ such that 
\begin{equation}
\label{eq:SDP_triggering2}
\left[ \begin{array}{ccc}
-\displaystyle \frac{\mu S \Omega S}{2} + 2 \sigma^2 I & \mu SX_1L & \mu S \Delta \\
\star & -I+\epsilon L^\top L & \mathbf{0} \\
\star& \star & -\epsilon I
\end{array} \right] \preceq 0 .
\end{equation}   
\end{enumerate}
Then, for any $\nu>0$, system (\ref{eq:closed-loop2})-(\ref{eq:triggering2}):
\begin{itemize}
\item[(a)] admits a global minimum inter-event time, in particular
for any solution $x$ with input disturbance $d$, the sequence of transmission times $\{t_k\}_{k\in\mathcal{N}}$ satisfies $t_{k+1}-t_k \geq \overline{\tau}(\sigma)$ for every $k \in \mathcal{N}$ where $\overline\tau(\sigma):=\frac{1}{\overline \alpha} \frac{\sigma}{1+\sigma}$,  $\overline\alpha:=\max\{\|X_1G\|+\|\Delta\| \|G\|,
\|X_1L\|+\|\Delta\| \|L\|,\sigma\delta / \nu\}$ and $\delta:=\|d\|_{\infty}$; 
\item[(b)] is practically exponentially ISS. \hfill $\Box$ 
\end{itemize}
\end{theorem}  

\emph{Proof.} We first prove item (a) of Theorem \ref{thm:practical}. We adapt for this purpose the  techniques in the  proof of \cite[Theorem IV.2]{Borgers-Heemels-tac14} carried out in the model-based setting. Let $x$ be a solution to system (\ref{eq:closed-loop2})-(\ref{eq:triggering2}) with input disturbance $d$. Let $\overline \nu:=\nu/\sigma$ and $k\in\mathcal{N}$. 
For any $t\in[t_k,t_{k+1})$
\begin{subequations}
\label{eq:Tab_MIET-practical-ISS}
\begin{alignat}{2}
& \frac{d}{dt} \frac{|e(t)|}{|x(t)|+\overline \nu} \nonumber \\
& \quad = \frac{e(t)^\top \dot e(t)}{|e(t)| (|x(t)|+\overline \nu)} - 
\frac{x(t)^\top \dot x(t) |e(t)|}{|x(t)| (|x(t)|+\overline \nu)^2} \\
& \quad \leq  \left( 1+ \frac{|e(t)|}{ |x(t)|+\overline \nu} \right) 
\frac{|\dot x(t)|}{|x(t)|+\overline \nu} \\
& \quad \leq \overline \alpha \left( 1+ \frac{|e(t)|}{ |x(t)|+\overline \nu} \right)^2 
\end{alignat}  
\end{subequations} 
with $\overline \alpha := \max\{\|X_1G\|+\|\Delta\| \|G\|,
\|X_1L\|+\|\Delta\| \|L\|,\delta /\overline \nu\}$ where $\delta:=\|d\|_{\infty}$.
The last inequality follows from the identities
$A+BK=(X_1-D_0)G$ and $BK=(X_1-D_0)L$, which imply
$\dot x = (X_1-D_0) G x + (X_1-D_0) Le + d$. This, in turn, gives 
$|\dot x| \leq \overline \alpha (|x|+|e|+\overline \nu)$.
Hence, the time needed for $|e|/(|x|+\overline \nu)$ to reach $\sigma$ 
(the event that triggers a new sampling) is not smaller than
the time denoted $\overline\tau(\sigma)$ needed for $\phi$ to reach $\sigma$, where $\phi$ is the solution to
the differential equation $\dot \phi = \overline \alpha (1+\phi)^2$ with initial value $\phi(0)=0$.
This time  is 
\begin{equation} \label{eq:bar_tau}
\overline \tau(\sigma) := \frac{1}{\overline \alpha} \frac{\sigma}{1+\sigma}.
\end{equation} 
Hence, $t_{k+1}-t_k \geq \overline \tau(\sigma)$, and item (a) of Theorem \ref{thm:practical} holds as $x$ and $k$ have been arbitrarily selected. We then derive that any solution to (\ref{eq:closed-loop2})-(\ref{eq:triggering2}) is complete.

Consider now item (b) of Theorem \ref{thm:practical}. By a Schur complement, \eqref{eq:SDP_triggering2} is equivalent to
\begin{eqnarray}
\label{eq:SDP_triggering2a}
\mu \begin{bmatrix}
- \displaystyle \frac{S \Omega S}{2} & SX_1L  \\
\star & \mathbf{0} \end{bmatrix} - \overline \Psi(\sigma) + \epsilon 
\begin{bmatrix} \mathbf{0} \\ L^\top \end{bmatrix} \begin{bmatrix} \mathbf{0} & L \end{bmatrix}   
\nonumber \\ + \epsilon^{-1} 
\begin{bmatrix} \mu S \Delta \\ \mathbf{0} \end{bmatrix} \begin{bmatrix} (\mu S \Delta)^\top & \mathbf{0} 
\end{bmatrix}  
\preceq 0 .
\end{eqnarray}
By applying  Lemma \ref{lem:Petersen_aux} in the Appendix with 
$B=[\begin{smallmatrix} \mathbf{0} \\ L^\top \end{smallmatrix}]$ and
$C=[\begin{smallmatrix} -\mu S \\ \mathbf{0} \end{smallmatrix}]^\top$ we obtain
\begin{eqnarray}
\label{eq:SDP_triggering2b}
\mu \begin{bmatrix}
- \displaystyle \frac{S \Omega S}{2} & SX_1L  \\
\star & \mathbf{0} \end{bmatrix} - \overline \Psi(\sigma) +  
\begin{bmatrix} \mathbf{0} \\ L^\top \end{bmatrix} D^\top 
\begin{bmatrix} -\mu S \\ \mathbf{0} \end{bmatrix}^\top 
\nonumber \\ +  
\begin{bmatrix} -\mu S \\ \mathbf{0} \end{bmatrix} D \begin{bmatrix} \mathbf{0} & L
\end{bmatrix}  
\preceq 0  \quad \forall D \in \mathcal D, 
\end{eqnarray}
which can be compactly written as $\mu \overline M(D) - \overline \Psi(\sigma) \preceq 0$ for any $D\in\mathcal{D}$.
Thus, if \eqref{eq:SDP_triggering2} feasible then
$\mu \overline M(D) - \overline \Psi(\sigma) \preceq 0$ for any $D\in\mathcal{D}$
and Assumption \ref{ass:D0} implies 
$\mu \overline M(D_0) - \overline \Psi(\sigma)\preceq 0$.
Let $x$ be a solution to system (\ref{eq:closed-loop2})-(\ref{eq:triggering2}) with input disturbance $d$ and $t\geq 0$. The triggering rule \eqref{eq:triggering2}
ensures $|e(t)|^2 \leq 2 \sigma^2 |x(t)|^2 + 2 \nu^2$. Therefore, $z(t)^\top \overline \Psi(\sigma) z(t) \leq 2 \nu^2$. Since $\mu \overline M(D_0) - \overline \Psi(\sigma)\preceq 0$
then  $z(t)^\top \overline M(D_0) z(t) \leq 2 \nu^2 / \mu$. Substituting this inequality into (\ref{eq:V_approx-last})
and resorting to standard Lyapunov arguments lead to the satisfaction of item (b) of Theorem \ref{thm:practical}. \hfill $\blacksquare$

\begin{figure*}[!t]
\caption{Results for Example \ref{ex:2}. Top figures report simulation results for
$\delta=0.1$ while bottom figures report the results for $\delta=0.5$.
Left: State trajectories.
Middle: Behavior of $|e(t)|$ and $\sigma |x(t)| + \nu$; a new sampling is 
triggered when $|e(t)|=\sigma |x(t)| + \nu$. 
Right: Behavior of the inter-sampling times $t_{k+1}-t_k$. 
The minimum inter-sampling time observed in simulation is $0.0184$ for 
$\delta=0.1$ and $0.0021$ for $\delta=0.5$, which is 
about one order of magnitude larger than the theoretical lower bound.
} \label{fig:ex2}
\hspace*{-0.2cm}
{\includegraphics[width=5cm]{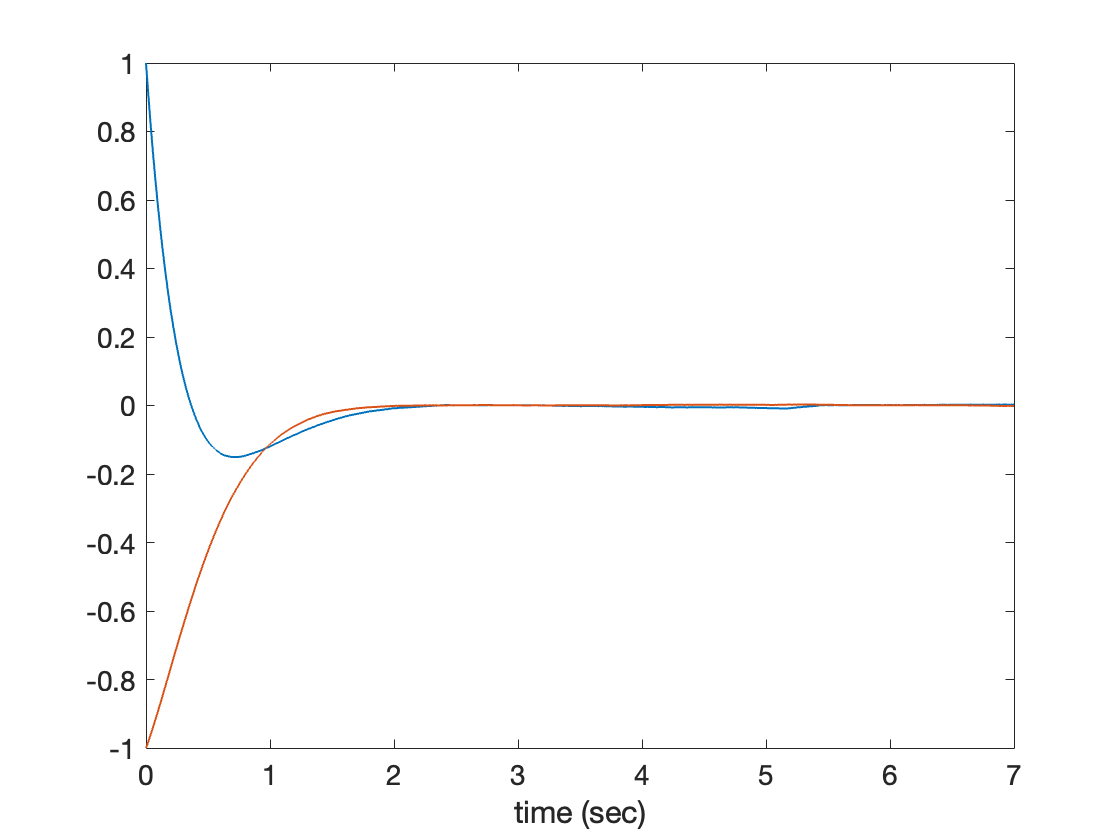}} \hspace*{-0.7cm}
{\includegraphics[width=5cm]{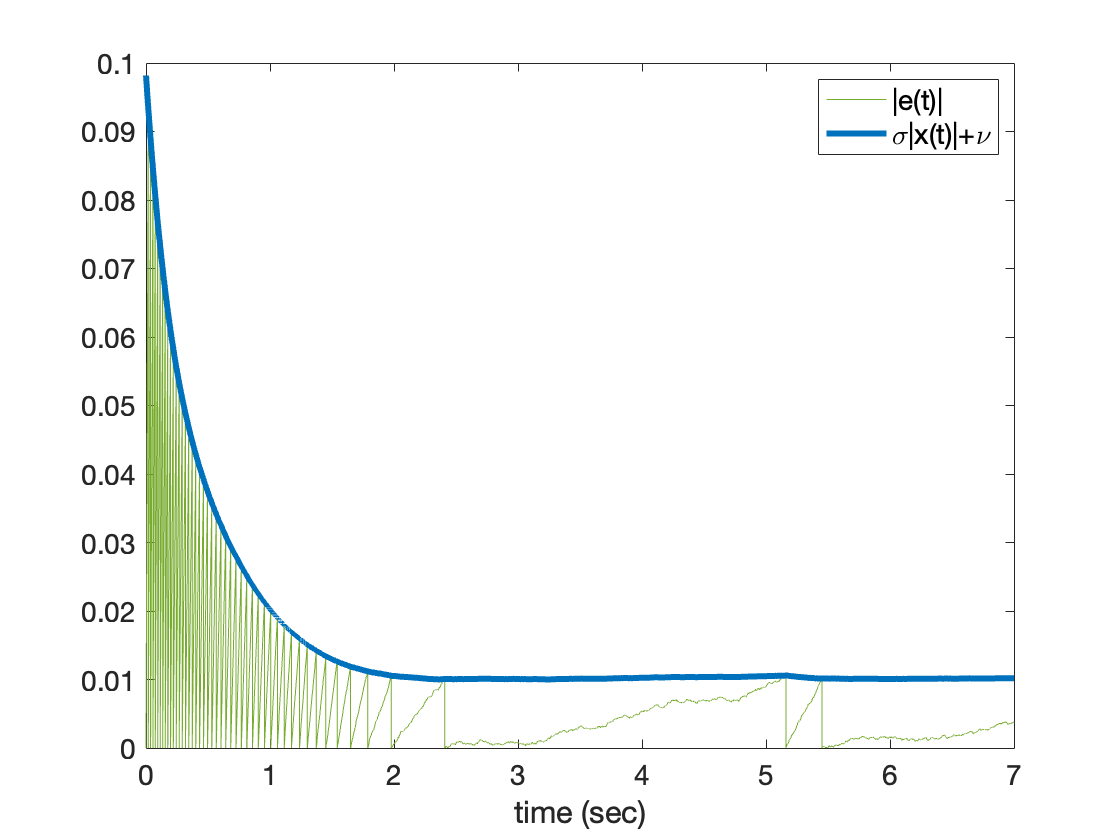}} \hspace*{-0.7cm} 
{\includegraphics[width=5cm]{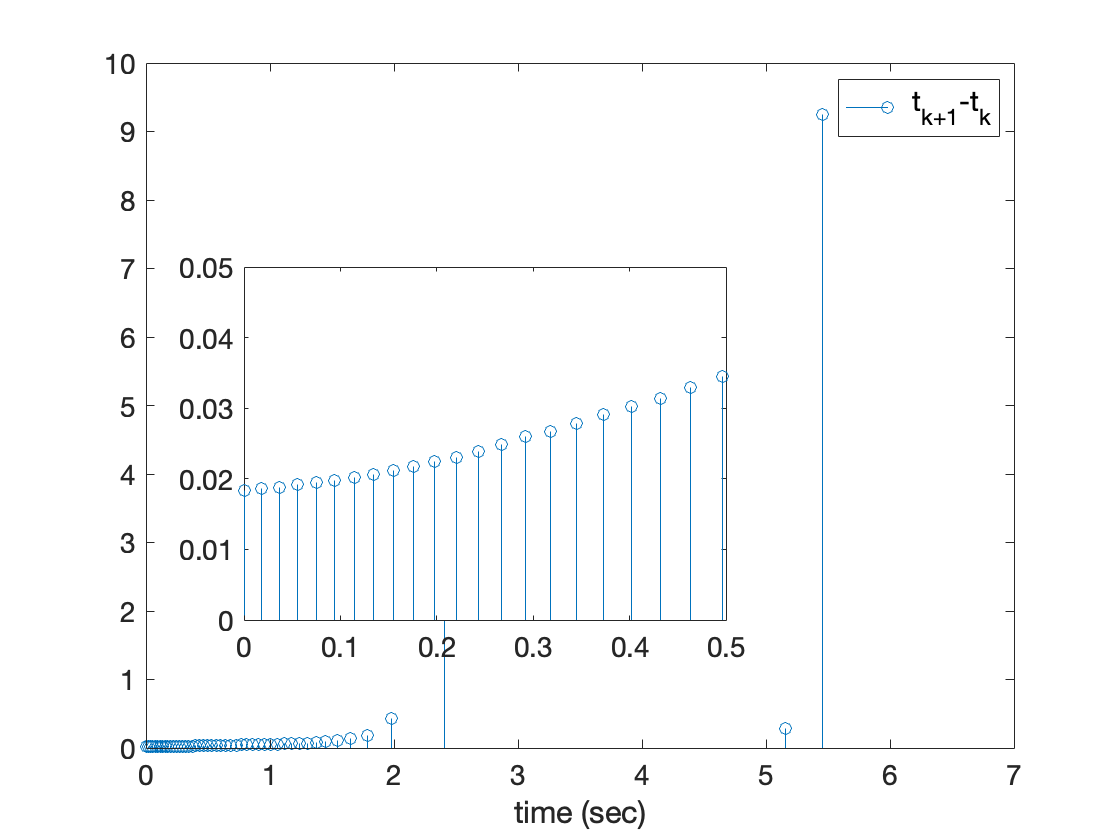}} \\
{\includegraphics[width=5cm]{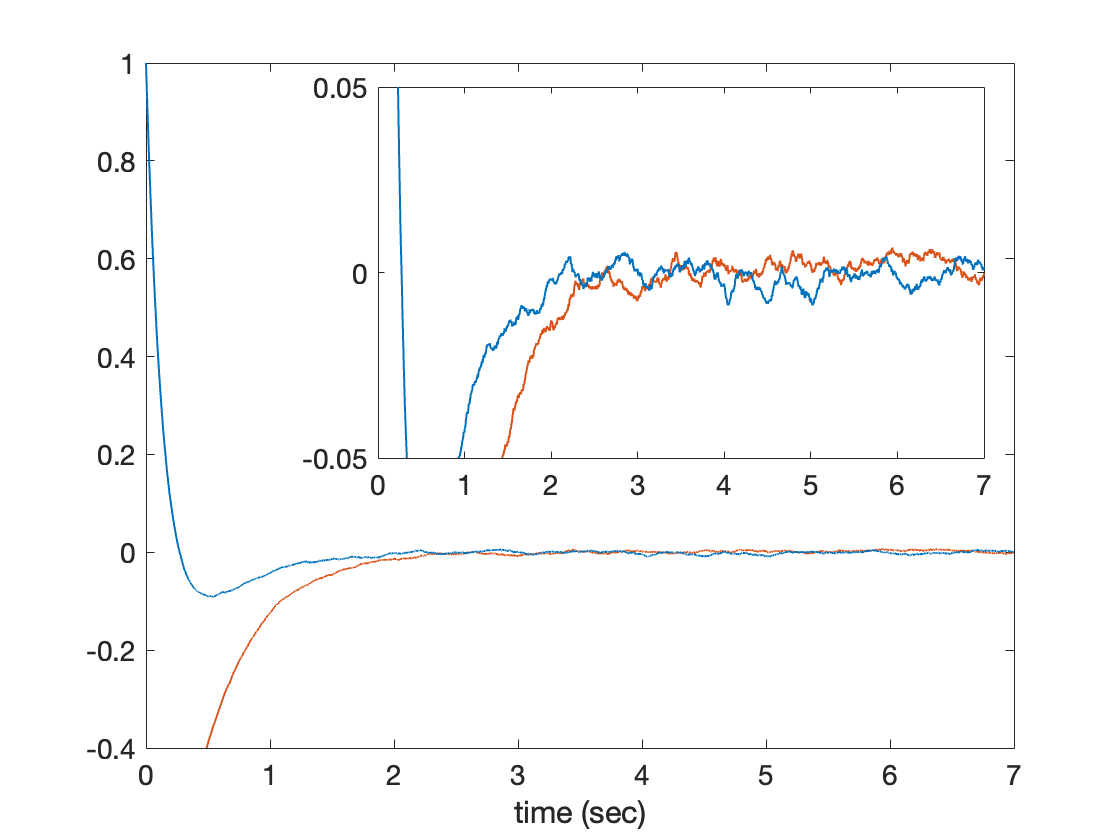}} \hspace*{-0.7cm}
{\includegraphics[width=5cm]{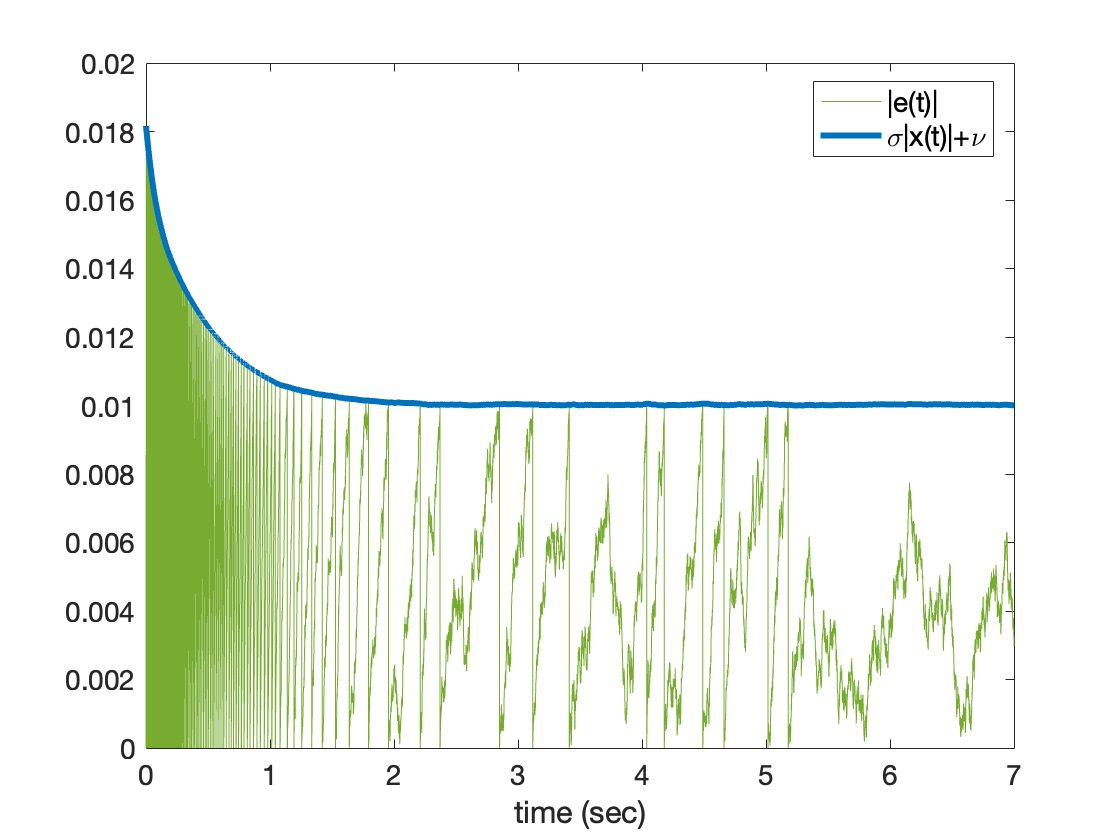}} \hspace*{-0.7cm} 
{\includegraphics[width=5cm]{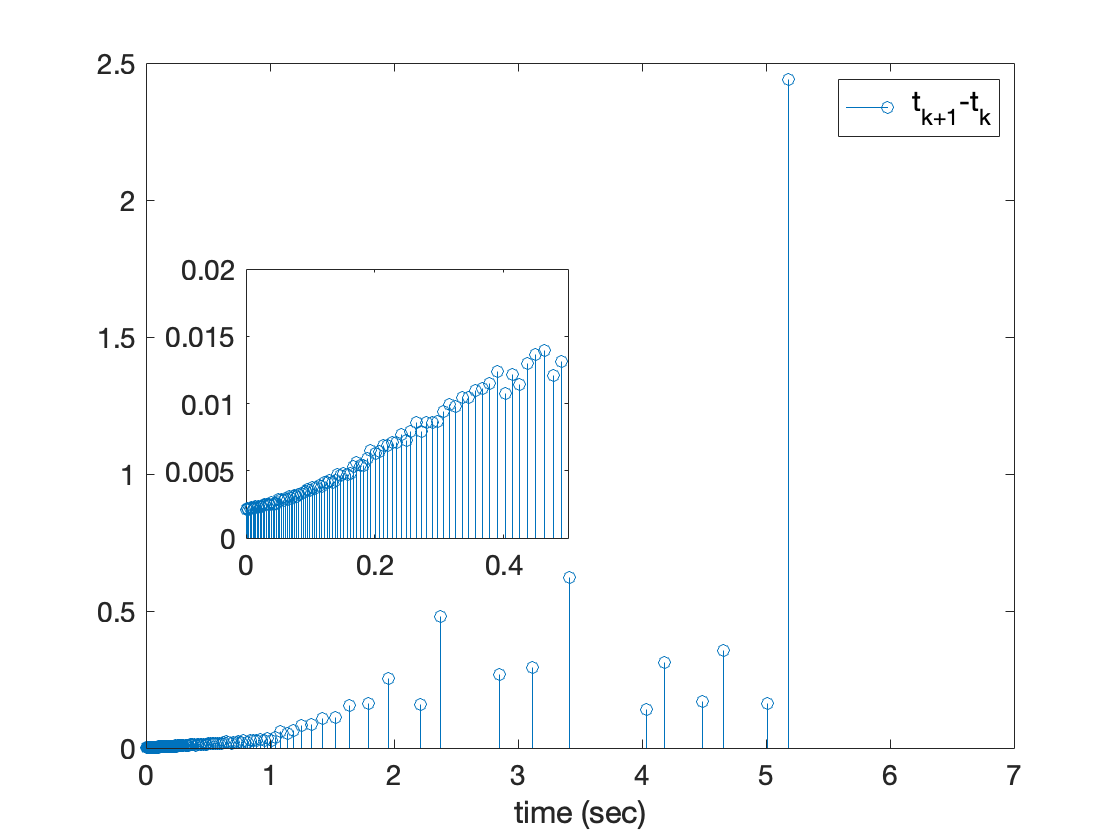}} 
\setcounter{equation}{\value{equation}}
\end{figure*}%

Similarly to $\sigma$, the choice of $\nu$ involves a trade off
between performance and the number of transmissions:
by decreasing $\nu$, the neighborhood of the origin to which the solutions to (\ref{eq:closed-loop2}), (\ref{eq:e-noisy}), (\ref{eq:triggering2})  converges to ``shrinks'' in view of item (ii) of Theorem \ref{thm:practical}, which typically leads to more transmissions. 
In particular, it follows from the proof of Theorem \ref{thm:practical} that, when $d\equiv 0$, the state converges to the ball of $\R^{n}$ centered at the origin of radius 
$\nu \sqrt{ \frac{2\kappa(S)}{\omega \mu} }$,
where $\kappa(S)$ is the condition number of $S$ and
$\omega$ is the smallest eigenvalue of $S\Omega S/2$. 
To ensure an exponential ISS property for the event-triggered controlled system, and not a practical exponential ISS property as in Theorem \ref{thm:practical}, a different triggering rule is needed (see Remark \ref{rem-special-cases} below): this is the purpose of the next section. 
Before that, we illustrate the results of Theorem \ref{thm:practical} for the system considered in Example \ref{ex:1}, and we provide two remarks on the triggering rule \eqref{eq:triggering2}.

\begin{example} \label{ex:2}
We consider the same system as in Example \ref{ex:1} but
this time a disturbance $d$ affects the system dynamics, with $\|d\|_{\infty} \leq \delta$.
As before, we collect the data by running an experiment with an input uniformly 
distributed in $[-1,1]$, and with initial state within the same interval. 
We consider two cases, $\delta=0.1$ and $\delta=0.5$. For both cases, we
first solve the SDP \eqref{eq:2SDP} with $\Omega:=10I$ to find $K$; all the choices of $\Omega$ that we tested in the form $\Omega=cI$, $c \in [1,100]$, led to $\overline \tau(\sigma)$ of similar magnitude. Then 
we solve the SDP \eqref{eq:SDP_triggering2} to find $\sigma$, in particular,
like \eqref{eq:SDP_triggering_max}, we cast \eqref{eq:SDP_triggering2}
as an optimization problem where we search for a solution maximizing 
$\sigma$. We use $\Delta=\delta\sqrt{T}I$ in both the SDPs. 
For $\delta=0.1$ we obtain 
$K = \begin{bmatrix} -3.1769  &  1.8145 \end{bmatrix}$
and $\sigma=0.0624$. Using the value $\nu=0.01$
in the triggering rule \eqref{eq:triggering2} gives 
the lower bound $\overline \tau(\sigma)=0.0135$ for the inter-sampling times. 
For $\delta=0.5$ we have 
$K = \begin{bmatrix} -6.8882  &  1.5924 \end{bmatrix}$
and $\sigma=0.0058$. Using $\nu=0.01$
in the triggering rule \eqref{eq:triggering2} gives
the lower bound $\overline \tau(\sigma)=3.9618e$--$04$.
For both the cases, we report simulation results in Figure \ref{fig:ex2}.

For both values of $\delta$, the sampling is very frequent initially
when the state is far from the origin, and becomes sporadic as soon
as the state gets smaller. Here, $\nu=0.01$ gives a good trade-off between 
performance and number of samplings. We note that $\nu$ can also 
be tuned \emph{online} since
its choice affects neither \eqref{eq:2SDP} nor \eqref{eq:SDP_triggering2},
hence it can be chosen after $K$ and $\sigma$ are determined.
A second remark regards $d$. The noise level 
affects the feasibility of both \eqref{eq:2SDP} and \eqref{eq:SDP_triggering2}, 
which become infeasible when the noise level becomes too
high. In this example, feasibility is preserved as long as the energy of the noise remains 
about half that of $u$ ($\delta \approx 0.5$, which corresponds to an input-disturbance 
signal-to-noise ratio of about $6$dB). Once we find a solution, however, 
practical exponential ISS holds \emph{irrespective} of the noise level.
Obviously, for large noise levels we might need to increase
$\nu$ in order to reduce the number of samplings. This is evident from
 \eqref{eq:bar_tau} where we see that increasing $\delta$ can make $\overline \alpha$
 larger, in which case $\overline \tau(\sigma)$ decreases.
\hfill $\blacksquare$
\end{example}

\begin{rem}\label{rem-mixed-triggering-actual-form} The proof of Theorem \ref{thm:practical} shows that we can consider this triggering rule $t_{k+1} = \inf \{t \in \mathbb R : t > t_k \text{ and } |e(t)|^2 = 2\sigma^2|x(t)|^2 + 2\nu^2 \}$ instead of (\ref{eq:triggering2}). This is clear from the stability analysis in the proof of  Theorem \ref{thm:practical}, and we also have that, for any given state, the next transmission time with this rule will occur not before the one generated by (\ref{eq:triggering2}) as $|e|\leq\sigma|x|+\nu$ implies $|e|^2\leq 2\sigma^2|x|^2+2\nu^2$ for any $(x,e)\in\R^{2n}$. As a result, $\overline\tau(\sigma)$ is also a global minimum inter-event time for the above policy. Note also that, since Theorem \ref{thm:practical} applies for any $\nu>0$,  in view of the definition of $\overline\Psi(\sigma)$ in (\ref{eq:Psi2}), we can equivalently consider the next triggering rule \begin{equation} \label{eq:mixed-triggering-rule-remark}
t_{k+1} = \inf \{t \in \mathbb R : t > t_k \text{ and } z(t)^\top\overline\Psi(\sigma)z(t) =  \nu\}.
\end{equation}
This formulation  will be convenient when addressing the generalized triggering conditions of  Section \ref{sect:other-triggering-rules}. \hfill $\Box$
\end{rem}

\begin{rem}\label{rem-special-cases} We discuss in this remark two special cases of \eqref{eq:triggering2}, which are not covered by Theorem \ref{thm:practical}. First, when $\nu=0$ in (\ref{eq:triggering2}),  
it is not possible to guarantee that the inter-sampling times remain strictly positive in general as shown in \cite{Borgers-Heemels-tac14}. 
Second, when $\sigma=0$ in (\ref{eq:triggering2}), we can also prove that system (\ref{eq:closed-loop2}), (\ref{eq:e-noisy}), (\ref{eq:triggering2}) is practically exponentially ISS for any $\nu>0$, as in item (b) of Theorem \ref{thm:practical}. However, in this case, the lower-bound on the inter-event times in item (a) of Theorem \ref{thm:practical} is no longer global but semiglobal, in the sense that it depends on the ball of initial conditions of the state (and the supremum of the disturbance on $[0,\infty)$) like in \cite[Theorem IV.4]{Borgers-Heemels-tac14}. In fact, for any $c_x>0$,  any solution $x$ to (\ref{eq:closed-loop2})-(\ref{eq:triggering2})  with $|x(0)|\leq c_x$ and input disturbance $d$,  any $k\in\mathcal{N}$ and almost all $t\in[t_k,t_{k+1}]$, 
$
\frac{d}{dt} |e(t)| \le c_\Phi |x(t)| + c_{{\rm e}} |e(t)| +\|d\|_\infty$, 
where $c_\Phi:=\|X_1G\| + \|\Delta\| \|G\|$ and $c_{{\rm e}}:=\|X_1 L\| + \|\Delta\| \|L\|$. By following similar lines as in the proof of Theorem \ref{thm:practical} there exists a constant $\overline{c}_x$, which depends on $c_x$, $\nu$ and $\|d\|_{\infty}$, such that 
$\frac{d}{dt} |e(t)| \le c_\Phi \overline c_x + c_{{\rm e}} |e(t)| +\|d\|_\infty$
from which we infer that
the time needed for $|e(t)|$ to grow from $0$ to $\nu$ is lower bounded by $\frac{1}{ c_{{\rm e}}}\ln \left(\frac{ c_{{\rm e}}\nu}{c_\Phi \overline c_x+\|d\|_\infty}+1\right)$, 
which indeed
depends on  the radius of the ball of initial conditions of the state $c_x$ and the supremum norm of $d$. Note that in this case, i.e., $\sigma=0$ and $\nu>0$, the design of triggering rule directly follows from Theorem \ref{thm:approx} as item (iii) of Theorem \ref{thm:practical} always holds when $\sigma=0$.  \hfill $\Box$
\end{rem}

\subsubsection{Time-regularized triggering condition}\label{subsubsect:time-reg}\label{sec:ISS} 
We propose in this section an alternative triggering condition to ensure an exponential ISS property, as opposed to practical exponential ISS as in Section \ref{subsubsect:mixed-triggering-rule}, at the price of potentially more transmissions as suggested by Example \ref{ex:3} provided hereafter.
 
Our starting point is the characterization of the
Lyapunov function given in \eqref{eq:V_approx}, i.e., for any $z=(x,e)\in\R^{2n}$ and $d\in\R^{n}$,
\begin{equation}
\label{eq:V_approx_ISS}
\begin{array}{l}
\left\langle\nabla V(x),(A+BK)x+BKe+d\right\rangle \\[0.1cm]
\quad \leq\, -x^\top S \Omega S x + 2 \left[  (X_1-D_0)L e + d \right]^\top S x \\[0.1cm]
\quad \leq\, -\omega_1 |x|^2 + \omega_2 |x||e| + \omega_3 |x| |d|,
\end{array} 
\end{equation}  
with $\omega_1$ the smallest eigenvalue of $S \Omega S$,
$\omega_2:=2 \| SX_1L\|+2 \| S \| \| \Delta \| \| L\|$ and  $\omega_3:=2\|S\|$. 
From the above expression it readily follows that a sufficient condition to ensure exponential ISS is to ensure along any solution to (\ref{eq:closed-loop2}), (\ref{eq:e-noisy}) and any $t$ in the domain of the solution 
\begin{equation}
\begin{array}{rll}\label{eq:bound_e}
|e(t)|  \leq  \sigma ( |x(t)| +  \|d\|_{[0,t]} ) 
\end{array}
\end{equation}
with 
\begin{equation}\label{eq:sigma-time-reg}
\begin{array}{rlll}
    \sigma  \in  \left(0,\omega_1/\omega_2\right),
\end{array}
\end{equation}
provided Zeno phenomenon does not occur. When (\ref{eq:bound_e}) holds, 
$\dot{V}(x(t)) \leq - (\omega_1 - \sigma\omega_2) |x(t)|^2
+ (\sigma\omega_2 +  \omega_3 ) |x(t)|\|d\|_{[0,t]}$ from which exponential ISS follows as $\omega_1 - \sigma\omega_2>0$  (again, provided Zeno phenomenon does not occur).
We first provide a \emph{model-based} condition that ensures \eqref{eq:bound_e}.
This result is a variant of \cite[Lemma 1]{dpt_DOS_2015}
in which we consider the logarithmic norm of $A$ instead of the induced $2$-norm
considered here. We consider the induced $2$-norm of $A$ because it is
somehow easier to infer from data\footnote{The result below holds for arbitrary $\sigma>0$. Imposing 
$\sigma< \omega_1/\omega_2$ is needed to make sure that \eqref{eq:bound_e} 
guarantees ISS.}.

\begin{lem} [\cite{dpt_DOS_2015}, Lemma 1] \label{lem:intersampling_ISS}
Let $K$ be  any feedback matrix that makes $A+BK$ Hurwitz and 
\begin{equation} \label{eq:tau_m}
\tau_m(\sigma) \!:=\! \left\{ 
\begin{array}{l} 
\displaystyle
\!\frac{1}{\|A\|} \log \left( \frac{\sigma}{1+\sigma}  
\frac{\|A\|}{\max\{\|A+BK\|,1\}} +1\right)\\  
\hfill \text{ if } A\neq0,  \\[0.2cm]
\!\displaystyle  \frac{\sigma}{1+\sigma}  
\frac{1}{\max\{\|A+BK\|,1\}}  \hfill \text{otherwise.}
\end{array}
\right.
\end{equation} 
Given any triggering policy and any solution $x$ to the corresponding closed-loop system (\ref{eq:closed-loop2}), (\ref{eq:e-noisy}) with input disturbance $d$,  for any $k\in\mathcal{N}$ and any $t\in[t_k,t_k+\tau_m(\sigma)]\cap[t_k,t_{k+1}]$,  $|e(t)| \leq \sigma ( |x(t)| +  \|d\|_{[0,t]} )$.
\hfill $\Box$
\end{lem}

The expression of $\tau_m(\sigma)$ in (\ref{eq:tau_m}) depends on the model via the terms $\|A\|$ and $\|A+BK\|$. A lower bound of $\tau_m(\sigma)$ can be derived based on the available data in $\mathbb{D}$. We use for this purpose data-based upper-bounds on $\|A\|$ and $\|A+BK\|$, as the expression in (\ref{eq:tau_m}) is monotonically decreasing in $\|A\|$ and $\|A+BK\|$.  
We notice for this purpose that $A+BK=(X_1-D_0)G$
with $G$ satisfying \eqref{eq:GK}, while $A$ satisfies
$[\begin{smallmatrix} B & A \end{smallmatrix} ]
= (X_1-D_0) [\begin{smallmatrix} U_0 \\ X_0 \end{smallmatrix}]^\dag$
where $M^\dag$ is the right inverse of the matrix $M$. Partitioning 
$[\begin{smallmatrix} U_0 \\ X_0 \end{smallmatrix}]^\dag = 
[\begin{smallmatrix} J_0 & V_0 \end{smallmatrix}]$ with $V_0$ having the same dimension 
as $A$, we thus have $A=(X_1-D_0)V_0$. 
Then, $\|A+BK\| \leq \|X_1G\| + \|\Delta\| \|G\|$ and 
$\|A\| \leq \|X_1V_0\| + \|\Delta\| \|V_0\|$, and the upper bounds are
both computable from data alone. We then have the next result.

\begin{lem} \label{lem:intersampling_ISS_data}
Let $K$ be any feedback matrix that makes 
$A+BK$ Hurwitz and 
\begin{equation} \label{eq:tau_d}
\tau_d(\sigma) := 
\frac{1}{c_A} \log \left( \frac{\sigma}{1+\sigma}  
\frac{c_A}{\max\{c_\Phi,1\}} +1\right)  
\end{equation} 
with $c_A:=\|X_1V_0\| + \|\Delta\| \|V_0\|$ and 
$c_\Phi=\|X_1G\| + \|\Delta\| \|G\|$.
Given any triggering policy and any solution $x$ to the corresponding closed-loop system (\ref{eq:closed-loop2}), (\ref{eq:e-noisy}) with input disturbance $d$,  for any $k\in\mathcal{N}$ and any $t\in[t_k,t_k+\tau_d(\sigma)]\cap[t_k,t_{k+1}]$,  $|e(t)| \leq \sigma ( |x(t)| +  \|d\|_{[0,t]} )$.\hfill $\Box$
\end{lem}

\emph{Proof.} The result follows from Lemma \ref{lem:intersampling_ISS}
and the fact that $\tau_d(\sigma) \leq \tau_m(\sigma)$. Specifically, for $A \neq0$ the inequality $\tau_d(\sigma) \leq \tau_m(\sigma)$
follows since $\tau_m(\sigma)$ decreases monotonically as $\|A\|$ and/or $\|A+BK\|$ increase,
while for $A = 0$ the inequality $\tau_d(\sigma) \leq \tau_m(\sigma)$ follows since
$\log(1+s) \leq s$ for every $s \geq 0$. \hfill $\blacksquare$ 
 
With this result in hands, we derive the next triggering policy: $t_0=0$ and 
\begin{equation} \label{eq:triggering_ISS}
t_{k+1} = 
\begin{array}{l}
\inf \left\{t\in\R\,:\, t \geq t_k+\tau_d(\sigma) \text{ and }  z(t)^\top \Psi(\sigma) z(t) \geq 0  \right\}  
\end{array}
\end{equation}  
with $\Psi(\sigma)$ and $\sigma$ as in  \eqref{eq:Psi}  and (\ref{eq:sigma-time-reg}), respectively. The rule in (\ref{eq:triggering_ISS}) is a time-regularized version of (\ref{eq:triggering}), which 
prevents arbitrary fast sampling by enforcing global minimum inter-event time $\tau_d(\sigma)>0$ given in (\ref{eq:tau_d}). We next  state the main result of this section. 
Afterwards, we discuss pros and cons of this approach with respect to the one presented in Section \ref{subsubsect:mixed-triggering-rule}.

\begin{theorem} \label{thm:ISS}
Suppose the following holds.
\begin{enumerate}
    \item[(i)] Assumptions \ref{ass:rich} and \ref{ass:D0} are verified with $\Delta$ given.
    \item[(ii)] Let $\Omega \succ 0$, SDP \eqref{eq:2SDP} is feasible and $K=U_0 Y (X_0 Y)^{-1}$ is the resulting controller as in Theorem \ref{thm:approx}.
\end{enumerate}
Let $\sigma$ and $\tau_d(\sigma)$ as in (\ref{eq:sigma-time-reg}) and \eqref{eq:tau_d}, respectively, then 
system \eqref{eq:closed-loop2}, \eqref{eq:e-noisy}, \eqref{eq:triggering_ISS} is exponentially ISS. \hfill $\Box$
\end{theorem}

\emph{Proof.} Let $x$ be a solution to \eqref{eq:closed-loop2}, \eqref{eq:e-noisy}, \eqref{eq:triggering_ISS} with input disturbance $d$. We first note that $x$ is complete as it cannot explode in finite time and there exists a strictly positive minimum inter-event time $\tau_d(\sigma)>0$, which excludes Zeno phenomenon. Let $k\in\mathcal{N}$. We know from Lemma \ref{lem:intersampling_ISS_data} that $|e(t)|\leq \sigma(|x(t)|+\|d\|_{[0,t]})$ for all $t\in[t_{k},t_k+\tau_d(\sigma)]\cap[t_k,t_{k+1}]=[t_k,t_k+\tau_d(\sigma)]$ here in view of \eqref{eq:triggering_ISS}. Moreover, if $t_{k+1}>\tau_d(\sigma)$, this means $|e(t)|\leq \sigma|x(t)|$ for any $t\in[t_k+\tau_d(\sigma),t_{k+1}]$ and thus $|e(t)|\leq \sigma(|x(t)|+\|d\|_{[0,t]})$. Therefore,  (\ref{eq:bound_e}) holds for $t\in[t_k,t_{k+1}]$. We derive that (\ref{eq:bound_e}) holds (with $\sigma$ in \eqref{eq:sigma-time-reg}) along $x$. Since $x$ and  $d$ have been chosen arbitrarily,  (\ref{eq:V_approx_ISS}) and (\ref{eq:bound_e}) hold along solutions and $V(x)$ is not affected by jumps at the triggering instants, we derive that system \eqref{eq:closed-loop2}, \eqref{eq:e-noisy}, \eqref{eq:triggering_ISS}  is exponentially ISS.\hfill $\blacksquare$ 



Compared with the triggering rule \eqref{eq:triggering2}, \eqref{eq:triggering_ISS} 
has the merit to guarantee exponential ISS rather than practical exponential  ISS under the conditions of Theorem \ref{thm:ISS}. Further,
this transmission policy  does not require to solve the SDP in \eqref{eq:SDP_triggering2},
thus removing the question of feasibility related to finding a sampling policy.
On the other hand, the next example shows that this new triggering rule may have certain disadvantages, as it may result in  many more transmissions.

\begin{figure*}[!t]
\caption{Results for Example \ref{ex:3}. Top figures report simulation results for
$\delta=0.1$ while bottom figures report the results for $\delta=0.5$.
Left: State trajectories.
Middle: Behavior of $|e(t)|$ and $\sigma |x(t)|$; a new sampling is 
triggered when $|e(t)|=\sigma |x(t)|$ if the next inter-sampling is not smaller than $\tau_d(\sigma)$
and after $\tau_d(\sigma)$ seconds otherwise. 
Right: Behavior of the inter-sampling times $t_{k+1}-t_k$. 
The minimum inter-sampling time observed in simulation is $0.0197$ for 
$\delta=0.1$ and $3.2511e$--$04$ for $\delta=0.5$, which is the same as the theoretical lower bound.
} \label{fig:ex3}
\hspace*{-0.2cm}
{\includegraphics[width=5cm]{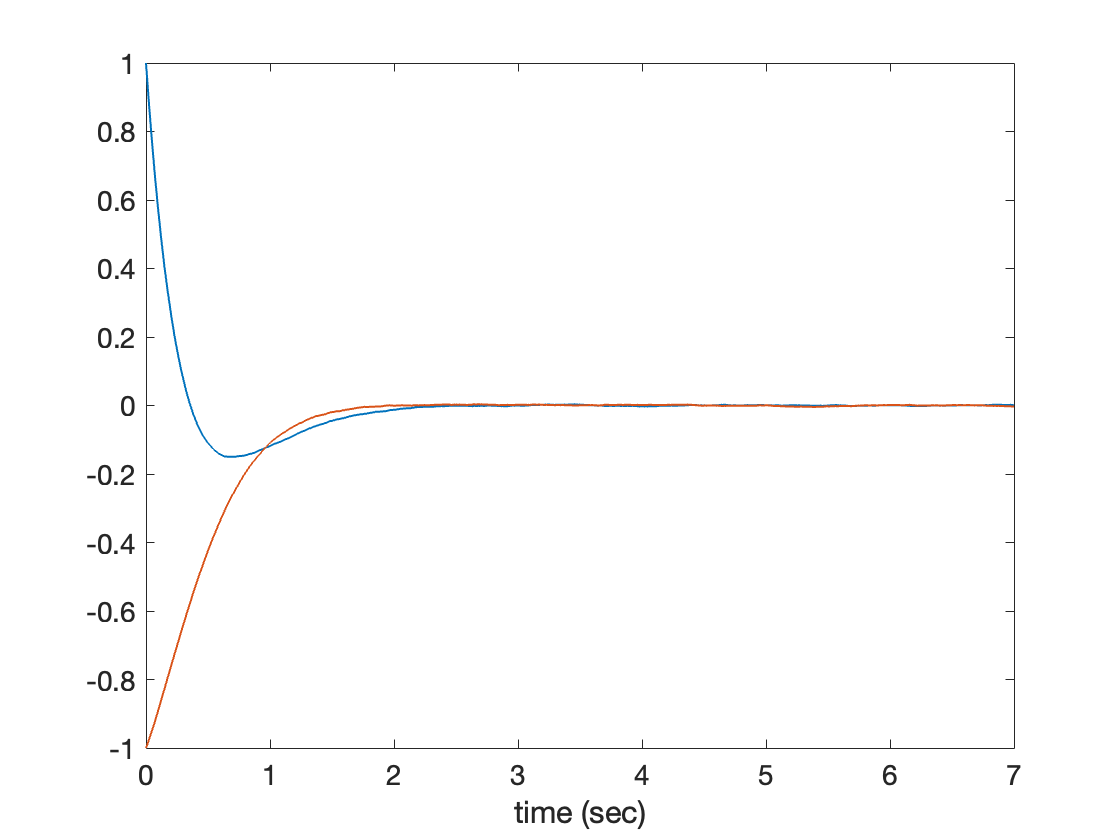}} \hspace*{-0.7cm}
{\includegraphics[width=5cm]{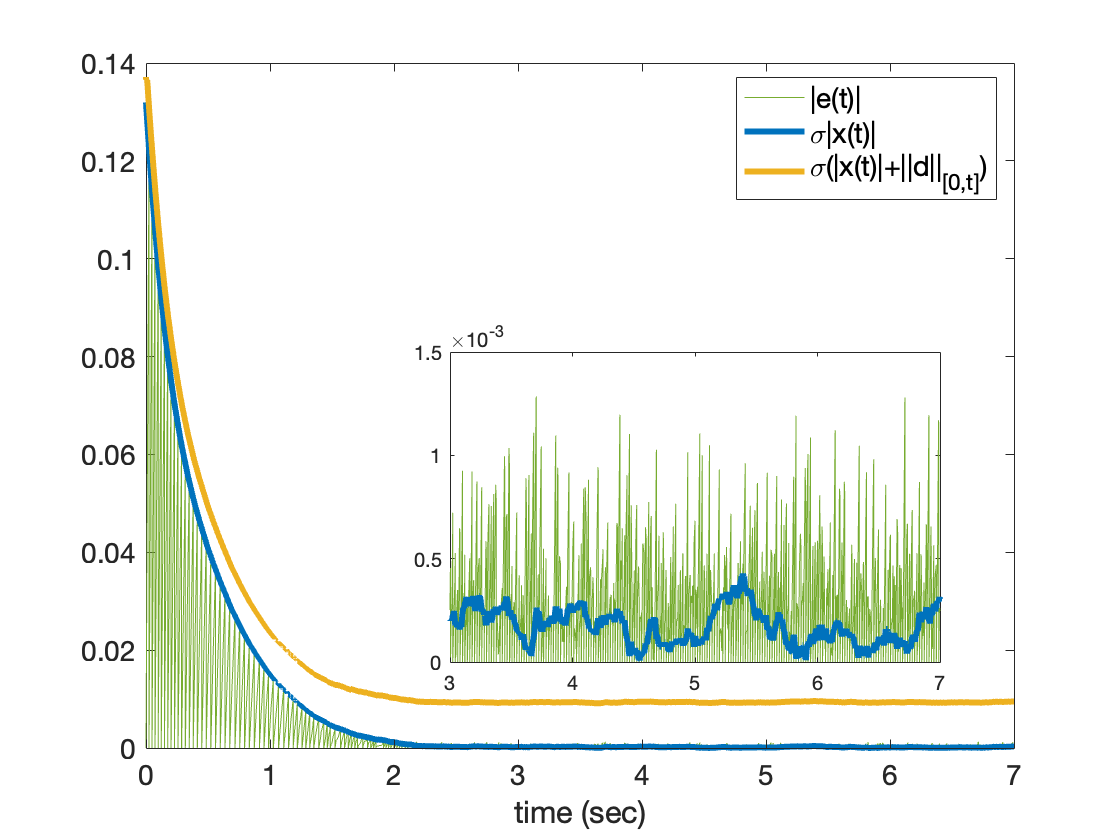}} \hspace*{-0.7cm} 
{\includegraphics[width=5cm]{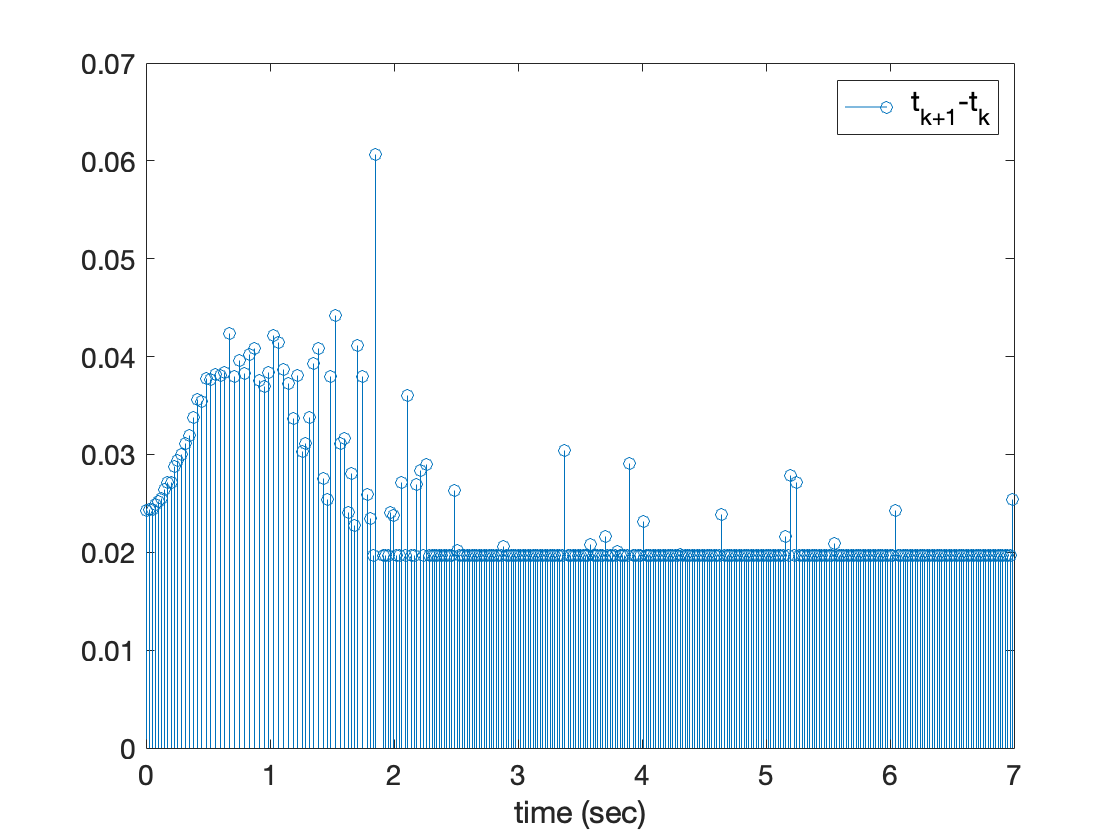}} \\
{\includegraphics[width=5cm]{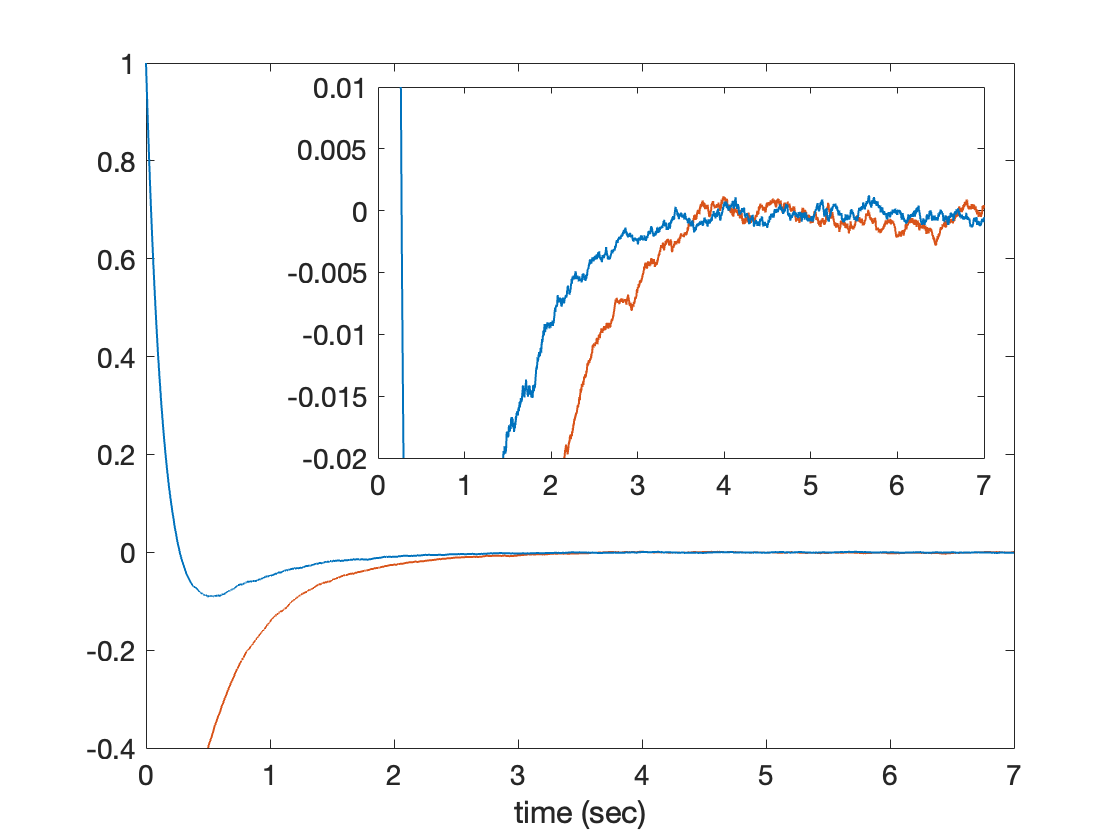}} \hspace*{-0.7cm}
{\includegraphics[width=5cm]{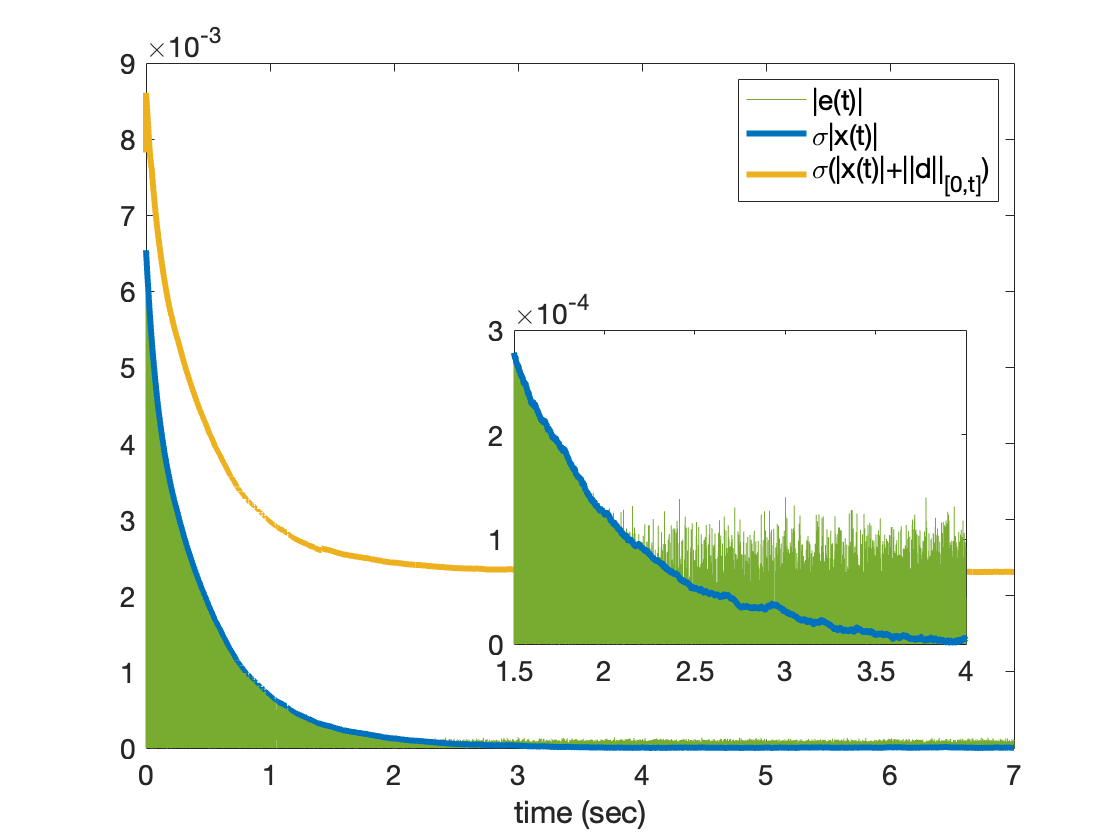}} \hspace*{-0.7cm} 
{\includegraphics[width=5cm]{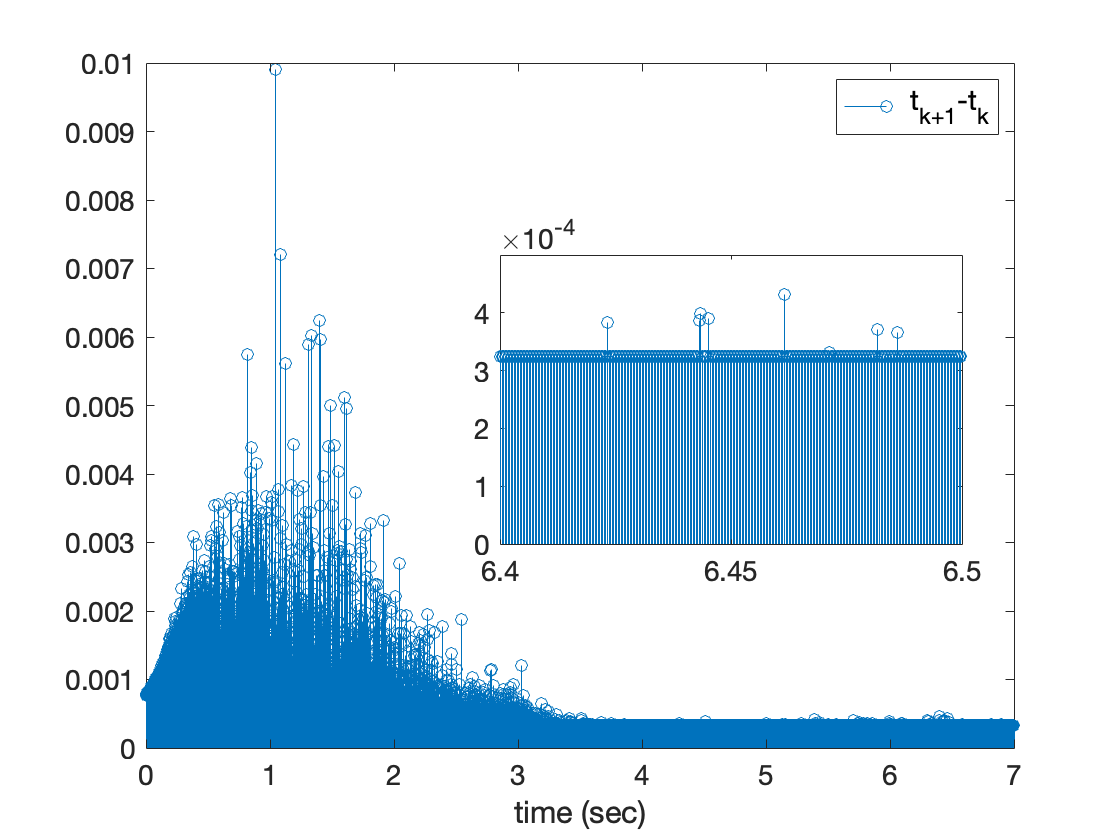}} 
\setcounter{equation}{\value{equation}}
\end{figure*}%

\begin{example} \label{ex:3}
We consider again Example \ref{ex:3} under the same setting.
As before, we consider two levels for the disturbance: $\delta=0.1$ and $\delta=0.5$. 
The control matrices are the same as before as 
the design of $K$ relies on \eqref{eq:2SDP} also in this case.
For $\delta=0.1$ a feasible value for $\sigma$ is $\sigma=0.0933$. Using this
value and the estimates $c_A$ and $c_\Phi$, we find 
$\tau_d(\sigma)=0.0197$. For $\delta=0.5$ a feasible value for $\sigma$ is 
$\sigma=0.0046$. Using this value and the estimates 
$c_A$ and $c_\Phi$ we find $\tau_d(\sigma)=3.2511e$--$04$.
We report simulation results in Figure \ref{fig:ex3}.

We see that the number of transmissions is much higher 
that the one obtained with the triggering rule \eqref{eq:triggering2}.
We see in particular that after an initial phase the triggering rule \eqref{eq:triggering_ISS}
eventually becomes almost periodic with $t_{k+1}=t_k+\tau_d(\sigma)$. 
Note that in Figure \ref{fig:ex3}, we can have $|e(t)|>\sigma|x(t)|$. This is fully consistent 
with the triggering rule \eqref{eq:triggering_ISS} which only guarantees 
$|e(t)|\leq \sigma(|x(t)|+\|d\|_{[0,t]})$ along solutions as shown in the proof of Theorem \ref{thm:ISS}. \hfill $\Box$
\end{example}

We conclude this section with a remark on the possible combination of \eqref{eq:triggering2} and \eqref{eq:triggering_ISS}, which will be useful in Section \ref{sect:other-triggering-rules}.

\begin{rem}\label{rem:space-time-reg-triggering} We can combine triggering rules \eqref{eq:triggering2} and \eqref{eq:triggering_ISS} as: $t_0=0$ and
\begin{equation} \label{eq:triggering-space-time-reg-prel}
\begin{array}{rll}
t_{k+1} = 
\inf \Big\{t \in\R \!\!&\!\! : \!\!&\!\!\!\!t\geq t_k+\tau_d(\sigma_1)  \\
&&\!\!\!\text{and } |e(t)| \geq \sigma_2 |x(t)|+\nu  \Big\},  
\end{array}
\end{equation}
or, in view of (\ref{eq:mixed-triggering-rule-remark}), 
\begin{equation} \label{eq:triggering-space-time-reg}
\begin{array}{rll}
t_{k+1} = 
\inf \Big\{t \in\R \!\!&\!\! : \!\!&\!\!\!\!t\geq t_k+\tau_d(\sigma_1) \\
& &\!\!\!\text{and } z(t)^\top \overline\Psi(\sigma_2) z(t)\geq \nu \Big\},  
\end{array}
\end{equation}
with $\sigma_1$ as in (\ref{eq:sigma-time-reg}) and $\sigma_2$ given by \eqref{eq:SDP_triggering2}. It is useful to distinguish parameters $\sigma_1$ and $\sigma_2$ used in  \eqref{eq:triggering2} and \eqref{eq:triggering_ISS}, respectively, as these may be assigned to different values. 
The existence of a global minimum inter-event time immediately follows from the fact that $\tau_d(\sigma_1)>0$. Moreover, in this case, the corresponding closed-loop system \eqref{eq:closed-loop2}, \eqref{eq:e-noisy}, \eqref{eq:triggering-space-time-reg} or \eqref{eq:mixed-triggering-rule-remark} is  practically exponentially ISS for $\nu>0$ under the conditions of Theorem \ref{thm:practical}; the proof of this property follows similar steps as those in the proofs of Theorems \ref{thm:practical} and \ref{thm:ISS}. \hfill $\Box$
\end{rem}

\section{Extension to other triggering rules}\label{sect:other-triggering-rules}

The approach presented in the previous section opens the door to the design of a range of robust data-based versions of existing (model-based) triggering rules. We illustrate this by presenting data-based versions of quadratic triggering policies \cite{Heemels2012} in Section \ref{subsect:quadratic-trig}, of dynamic event-triggered control \cite{Girard-tac15,Dolk-et-al-tac17} in Section \ref{subsect:dynamic-etc}, and of the technique, which consists in imposing a desired (typically decreasing) threshold on $V$, like in  \cite{Wang-Lemmon-aut11,Mazo-Anta-Tabuada-Aut10}, in Section \ref{subsect:decreasing-V}. 



\subsection{Quadratic policies}\label{subsect:quadratic-trig}

The triggering rule designed for the noise-free case in Section \ref{subsect:trig-rule-noisefree}, which has then been ``robustified'' in Section \ref{sec:noise}, is a special type of quadratic policies \cite{Heemels2012}. We can extend these results to triggering policies based on more general quadratic forms, which may be advantageous to reduce the number of transmissions. We first explain how this can be done in the noise-free case.

The triggering rule in  \eqref{eq:triggering} developed when $d\equiv 0$ in \eqref{system} is a special type of the general quadratic policy: $t_0=0$ and \begin{equation} \label{eq:triggering-quadratic}
t_{k+1} = \left\{ 
\begin{array}{l}
\inf \{t \in \mathbb R : t > t_k \text{ and } z(t)^\top \widetilde\Psi z(t) =0  \}  \\[0.1cm]
\hfill \text{if } x(t_k)\neq0,  \\[0.2cm]
 + \infty \hfill \text{otherwise,}
\end{array}
\right.
\end{equation}
with $\widetilde \Psi\in\R^{2n\times 2n}$ symmetric to be designed. In Section \ref{subsect:trig-rule-noisefree}, we focused on the special case where $\widetilde \Psi=\Psi(\sigma)$ and $\Psi(\sigma)$ as in (\ref{eq:Psi}) but other choices of $\widetilde\Psi$ are possible as suggested by the model-based results in, e.g.,  \cite{Heemels2012,Tabuada07}. To synthesize $\widetilde \Psi\in\R^{2n\times 2n}$ in (\ref{eq:triggering-quadratic}), we follow a similar approach as in Theorem \ref{thm:exact_sampling}, as formalized next.

\begin{proposition}\label{prop:quadratic-triggering-rule-noise-free} Suppose Assumption   \ref{ass:rich} holds and consider system (\ref{eq:closed-loop}), (\ref{eq:e}), (\ref{eq:triggering-quadratic})  
with $K=U_0Y(X_0Y)^{-1}$, $Y$ being any solution to \eqref{eq:SDP}. Let $\widetilde\Psi\in\R^{2n\times 2n}$ symmetric, $\mu,\,\sigma>0$ be such that the next SDP (in the decision variables 
$\mu$, $\sigma^2$ and $\widetilde\Psi$) is satisfied 
{\setlength\arraycolsep{2pt} 
\begin{equation}
\begin{array}{rllll}
\label{eq:SDP_triggering-quadratic}
\mu M - \widetilde{\Psi} & \prec & 0 \\
\widetilde\Psi & \preceq & \Psi(\sigma),
\end{array}
\end{equation}}%
with $\Psi(\sigma)$ and $M$ as in (\ref{eq:Psi}) and  (\ref{eq:V_exact}), respectively, $S=(X_0Y)^{-1}$, and $L$ as in \eqref{eq:L}. 
Then the following holds.
\begin{itemize}
\item[(a)] The system admits $\tau(\sigma)$ defined in item (a) of Theorem \ref{thm:exact_sampling} as global minimum inter-event time with $\sigma$ as in \eqref{eq:SDP_triggering-quadratic}.
\item[(b)] The origin of the system is  globally exponentially stable.
\hfill $\Box$
\end{itemize}  
\end{proposition}

\emph{Proof.} We first note that (\ref{eq:SDP_triggering-quadratic}) is feasible as it is satisfied with $\widetilde\Psi=\Psi(\sigma)$ for some small enough $\sigma>0$ according to the proof of Theorem \ref{thm:exact_sampling}. 

Because of the second inequality in (\ref{eq:SDP_triggering-quadratic}), given any solution $x$ to system (\ref{eq:closed-loop}), (\ref{eq:e}), (\ref{eq:triggering-quadratic}), at any transmission time $t_k$ with $k\in\mathcal{N}$, $t_{k+1}-t_k$ is greater than or equal the time it takes for $z^\top\Psi(\sigma)z$ to grow from $(x(t_k),0)^\top \Psi(\sigma)(x(t_k),0)$ to $0$, which corresponds to the inter-transmission time we would obtain with (\ref{eq:triggering}) starting at $x(t_k)$. Item (a) of Theorem \ref{thm:exact_sampling} ensures  the latter is lower bounded by $\tau(\sigma)$ in (\ref{eq:tau}). As a consequence, $t_{k+1}-t_k\geq \tau(\sigma)$ and since we have taken an arbitrary solution $x$ and an arbitrary $k\in\mathcal{N}$,  item (a) of Proposition \ref{prop:quadratic-triggering-rule-noise-free} holds. The proof of item (b) of Proposition \ref{prop:quadratic-triggering-rule-noise-free} follows the same steps as the proof of item (b) of Theorem \ref{thm:exact_sampling} in view of the first inequality in (\ref{eq:SDP_triggering-quadratic}). \hfill $\blacksquare$

The conditions in Proposition \ref{prop:quadratic-triggering-rule-noise-free} provide more flexibility in the design of matrix $\widetilde\Psi$ compared to Theorem \ref{thm:exact_sampling}, which becomes a particular case by taking $\widetilde\Psi=\Psi(\sigma)$ and $\Psi(\sigma)$ as in \eqref{eq:Psi}.

For the noisy case, to present the results in a compact way, we take inspiration from Remark \ref{rem:space-time-reg-triggering}, in particular \eqref{eq:mixed-triggering-rule-remark}, and consider the next triggering rule, which allows combining the techniques of Sections \ref{subsubsect:mixed-triggering-rule} and \ref{subsubsect:time-reg}  in a unified manner: $t_0=0$ and
\begin{equation} \label{eq:triggering-quadratic-noisy}
\begin{array}{l}
t_{k+1} = 
\inf \Big\{t \in\R \,:\, t\geq t_k+\overline{\tau}_d(\sigma_1) \text{ and }  z(t)^{\top}\widetilde{\Psi} z(t)\geq \nu \Big\}
\end{array}
\end{equation}  
where $\sigma_1$ is as in (\ref{eq:sigma-time-reg}), $\overline{\tau}_d(\sigma_1)\in\{0,\tau_d(\sigma_1)\}$ with $\tau_d(\sigma_1)$ in (\ref{eq:tau_d}), $\nu\geq 0$ arbitrary, and  $\widetilde{\Psi}\in\R^{2n\times 2n}$ symmetric to be designed. We have the next result. 

\begin{proposition}\label{prop:quadratic-triggering-rule-noisy} Suppose the following holds.
\begin{enumerate}
    \item[(i)] Assumptions \ref{ass:rich} and \ref{ass:D0} are verified with $\Delta$ given.
    \item[(ii)] Let $\Omega \succ 0$, SDP \eqref{eq:2SDP} is feasible and $K=U_0Y(X_0Y)^{-1}$ is the resulting controller as in Theorem \ref{thm:approx}.
    \item[(iii)] There exist $\mu, \epsilon, \sigma_2 > 0$ and $\widetilde\Psi\in\R^{2n\times 2n}$ symmetric such that
    {\setlength\arraycolsep{2pt} 
    \begin{equation}
    \begin{array}{rllll}\label{eq:SDP_triggering-quadratic-noisy}
    \left[ \begin{array}{cc|c}
    -\displaystyle \frac{\mu S \Omega S}{2}  & \mu SX_1L & \mu S \Delta \\
    \star & \epsilon L^\top L & \mathbf{0} \\
    \hline
    \star& \star & -\epsilon I
    \end{array} \right]  & \preceq &
    \left[ \begin{array}{c|c}
    \widetilde\Psi & \mathbf{0}\\
    \hline
    \mathbf{0} & \mathbf{0}
    \end{array} \right]
    \\
    \widetilde\Psi & \preceq & \overline{\Psi}(\sigma_2),
    \end{array}\end{equation}}%
\end{enumerate}
with
$L$,  $\Delta$ and  $\overline\Psi(\sigma_2)$ 
as in \eqref{eq:L}, \eqref{eq:noise_model} and \eqref{eq:Psi2}, respectively.
If $\overline{\tau}_d(\sigma_1)=\tau_d(\sigma_1)$ with $\sigma_1$ as in (\ref{eq:sigma-time-reg}) or $\nu>0$ in \eqref{eq:triggering-quadratic-noisy},  system  \eqref{eq:closed-loop2}, \eqref{eq:e-noisy}, \eqref{eq:triggering-quadratic-noisy}:
\begin{enumerate}
    \item[(a)] admits global inter-event time $\max\{\overline{\tau}_d(\sigma_1),\widetilde\tau(\sigma_2)\}$ with $\widetilde\tau(\sigma_2)=\overline\tau(\sigma_2)$ as in item (a) of Theorem \ref{thm:practical} when $\nu>0$, and $\widetilde\tau(\sigma_2)=0$ when $\nu=0$;
    \item[(b)] is  practically exponentially ISS when $\nu>0$, and is exponentially ISS when $\nu=0$. \hfill $\Box$
\end{enumerate}
\end{proposition}

\emph{Proof.} Item (a) of Proposition \ref{prop:quadratic-triggering-rule-noisy} is immediate when $\overline{\tau}_d(\sigma_1)=\tau_d(\sigma_1)>0$ and $\nu=0$ in view of \eqref{eq:triggering-quadratic-noisy}. When $\overline{\tau}_d(\sigma_1)=0$, then $\nu>0$. In this case, we have that for a given state at which a transmission occurs, the next transmission will occur not earlier than the one generated by triggering rule \eqref{eq:mixed-triggering-rule-remark} in Remark \ref{rem-mixed-triggering-actual-form} as $\widetilde\Psi\preceq\overline\Psi(\sigma_2)$, see  \eqref{eq:SDP_triggering-quadratic-noisy}. Since system  \eqref{eq:closed-loop2}, \eqref{eq:e-noisy}, \eqref{eq:mixed-triggering-rule-remark} admits global inter-event time $\overline\tau(\sigma_2)$ according to item (a) of Theorem \ref{thm:practical}, so does system \eqref{eq:closed-loop2}, \eqref{eq:e-noisy}, \eqref{eq:triggering-quadratic-noisy} and item (a) of Proposition \ref{prop:quadratic-triggering-rule-noisy} holds in this case. The last case where $\overline\tau_d(\sigma_1)=\tau_d(\sigma_1)$ and $\nu>0$ similarly follows.

To prove item (b) of Proposition \ref{prop:quadratic-triggering-rule-noisy}, we consider a solution $x$ to \eqref{eq:closed-loop2}, \eqref{eq:e-noisy}, \eqref{eq:triggering-quadratic-noisy} with input disturbance $d$ and $k\in\mathcal{N}$. Let $t\in[t_k,t_{k+1})$. If $t\leq t_k+\overline{\tau}_d(\sigma_1)$, then (\ref{eq:bound_e}) holds by Lemma \ref{lem:intersampling_ISS_data} and we have from (\ref{eq:V_approx_ISS})
%
%
%
{\setlength\arraycolsep{2pt} 
\begin{equation}
\begin{array}{rllll}\label{eq:proof-prop-quad-noisy-lyap-ineq-before-taud}   \dot V(x(t)) & \leq & -(\omega_1-\sigma_1\omega_2)|x(t)|^2 \\
& & +(\sigma_1\omega_2+\omega_3)|x(t)|\|d\|_{[0,t]},
\end{array}    
\end{equation}}%
with $\omega_1-\sigma_1\omega_2>0$ as $\sigma_1$ is such that (\ref{eq:sigma-time-reg}) holds. If $t\in[t_{k}+\overline{\tau}_d(\sigma_1),t_{k+1}]$, then $z(t)^\top\widetilde{\Psi}z(t)\leq \nu$ according to \eqref{eq:triggering-quadratic-noisy}. We deduce that $\mu\overline M(D_0)-\widetilde\Psi\preceq 0$ from the first inequality in (\ref{eq:SDP_triggering-quadratic-noisy}) and by following similar developments as in \eqref{eq:SDP_triggering2a} and \eqref{eq:SDP_triggering2b}. Consequently, since $z(t)^\top\widetilde{\Psi}z(t)\leq \nu$, $\mu z(t)^\top\overline M(D_0)z(t)\leq \nu$ and we have from (\ref{eq:V_approx}d)
\begin{equation}\label{eq:proof-prop-quad-noisy-lyap-ineq-after-taud} 
\dot{V}(x(t))\leq - x(t)^\top \left( \frac{S \Omega S}{2} \right) x(t) + \nu/\mu + 2 d(t)^\top S x(t),
\end{equation}
where we recall that $S\Omega S\succ 0$.

Based on \eqref{eq:proof-prop-quad-noisy-lyap-ineq-before-taud} and \eqref{eq:proof-prop-quad-noisy-lyap-ineq-after-taud}, we deduce that system  \eqref{eq:closed-loop2}, \eqref{eq:e-noisy}, \eqref{eq:triggering-quadratic-noisy} is  practically exponentially ISS; recall that $V(x)$ is not affected by jumps at the triggering instants. Furthermore,  when $\nu=0$, system  \eqref{eq:closed-loop2}, \eqref{eq:e-noisy}, \eqref{eq:triggering-quadratic-noisy} is exponentially ISS. \hfill $\blacksquare$

Like before, Proposition \ref{prop:quadratic-triggering-rule-noisy} extends Theorems \ref{thm:practical} and \ref{thm:ISS} to more general quadratic triggering rules.

\begin{rem} In Proposition \ref{prop:quadratic-triggering-rule-noisy}, consistently with Remark \ref{rem-mixed-triggering-actual-form}, there are parameters $\sigma_1$, used to design $\tau_d(\sigma_1)$, and $\sigma_2$, used to synthesize $\widetilde\Psi$. We can obviously take a single parameter $\sigma$ by selecting the minimum value between feasible $\sigma_1,\sigma_2$. 
\hfill $\Box$    
\end{rem}

\subsection{Dynamic event-triggered control}\label{subsect:dynamic-etc}

The triggering rules considered so far are static, in the sense that they rely on algebraic conditions involving  $x$ and $e$ (and possibly the elapsed time since the last transmission). We can also add auxiliary variables to define the triggering rules as advocated in e.g., \cite{Postoyan2014,Girard-tac15}, in order to potentially further reduce the number of transmissions while preserving stability. In this section, we show how the results of Section \ref{subsect:quadratic-trig} can be extended to dynamic triggering rules like in \cite{Girard-tac15,Dolk-et-al-tac17} where model-based results are presented.

Consistently with the structure of the paper so far, we first focus on the noise-free case, before taking into account  $d$ in \eqref{system} again. We introduce for this purpose variable $\eta\in\R_{\geq 0}$, whose dynamics is, for any $t\in\R_{\geq 0}$,
\begin{equation}
\begin{array}{rlll}    
\dot \eta(t)    = -\lambda \eta(t) - z(t)^\top\widetilde\Psi z(t), & \eta(0)\geq 0,
\end{array}\label{eq:eta-noise-free}
\end{equation}
where $\lambda>0$ is arbitrary and $\widetilde{\Psi}\in\R^{2n\times 2n}$ is symmetric and has to be designed. The purpose of variable $\eta$ is to filter the quadratic term considered in \eqref{eq:triggering-quadratic} in Section \ref{subsect:quadratic-trig} (in the noise-free case). Variable $\eta$ does not experience jumps at transmission instants. Note that the state vector is now $(x,\eta)$ and we will establish that the origin is globally asymptotically stable for the augmented system, as formalized in Proposition \ref{prop:dynamic-triggering-rule-noise-free} below and discussed afterwards. 

The triggering rule is defined as\footnote{\label{footnote:dynamic-etc-theta=0}When $\theta=0$, triggering instants are only allowed when $\eta(t)\leq 0$ \emph{and} $\dot{\eta}(t)\leq 0$, which can be computed on-line in the noise-free case. We do not specify this condition in \eqref{eq:triggering-dynamic-noise-free} to not overload the corresponding equation. Note that this extra condition, namely $\dot\eta(t)\leq 0$ when $\theta=0$, will not be needed when doing time-regularization in the noisy case in \eqref{eq:triggering-dynamic-noisy}.}: $t_0=0$ and
\begin{equation} \label{eq:triggering-dynamic-noise-free}
t_{k+1} = \left\{ 
\begin{array}{l}
\inf \Big\{t \in\R\,:\,t> t_k \text{ and } 
 \eta(t)-\theta z(t)^\top \widetilde{\Psi} z(t) \leq 0  \Big\}  \\[0.1cm]
\hfill \text{if } (x(t_k),\eta(t_k))\neq0,  \\[0.2cm]
 + \infty \hfill \text{otherwise,}
\end{array}
\right.
\end{equation}
where  $\theta\in\R_{\geq 0}$ is an additional arbitrary tuning parameter. We have the next result.

\begin{proposition} \label{prop:dynamic-triggering-rule-noise-free} Suppose the conditions of Proposition \ref{prop:quadratic-triggering-rule-noise-free} are satisfied. Then system (\ref{eq:closed-loop}), (\ref{eq:e}), (\ref{eq:eta-noise-free}), (\ref{eq:triggering-dynamic-noise-free}): 
\begin{itemize}
\item[(a)] admits $\tau(\sigma)$ in item (a) of Theorem \ref{thm:exact_sampling} as a global minimum inter-event time with $\sigma$ as in Proposition \ref{prop:quadratic-triggering-rule-noise-free};
\item[(b)] is globally asymptotically stable, in particular  there exist $c_1\geq 1$ and $c_2>0$ such that any solution  $(x,\eta)$ satisfies $|(x(t),\sqrt{\eta(t)})|\leq c_1 e^{-c_2 t}|(x(0),\sqrt{\eta(0)})|$ for any $t\geq 0$. 
\hfill $\Box$
\end{itemize}  
\end{proposition}

\emph{Proof.} Let $(x,\eta)$ be a solution to the corresponding system  (\ref{eq:closed-loop}), (\ref{eq:e}), (\ref{eq:eta-noise-free}), (\ref{eq:triggering-dynamic-noise-free}). We have that $\eta(t)\geq 0$ for any $t$ in the domain of the solution by following the same reasoning as in the proof of \cite[Lemma 2.2]{Girard-tac15}. We also have that, given any $k\in\mathcal{N}$, the next transmission will occur not earlier than the one generated by triggering rule \eqref{eq:triggering-quadratic}  by  invoking similar arguments as in the proof of \cite[Proposition 2.3]{Girard-tac15}. Consequently, item (a) of Proposition \ref{prop:dynamic-triggering-rule-noise-free} holds in view of item (a) of Proposition \ref{prop:quadratic-triggering-rule-noise-free}. 

Let $z=(x,e)\in\R^{2n}$ and $\eta\in\R_{\geq 0}$. We define $U(x,\eta)=V(x)+\eta/\mu$ with $V(x)=x^\top S x$ for any $x\in\R^{n}$ as in Section \ref{subsect:trig-rule-noisefree} and $\mu$ as in \eqref{eq:SDP_triggering-quadratic}. We have from the latter inequality that there exists $\varepsilon>0$ independent of $z$ and $\eta$ such that
{\setlength\arraycolsep{2pt} 
\begin{equation}
\begin{array}{rlllll}
\left\langle \nabla U(x,\eta),\left(\dot x,\dot\eta\right)\right\rangle & = & z^{\top} M z-\lambda\eta/\mu -1/\mu \, z^\top \widetilde\Psi z \\
& \leq & -\varepsilon |z|^{2} - \lambda \eta/\mu,
\end{array}\label{eq:proof-prop-dynamic-triggering-lyap-noise-free}
\end{equation}}%
where we write with some abuse of notation $\dot x = (A+BK)x+BKe$ and $\dot\eta=-\lambda \eta - z^\top\widetilde\Psi z$. By integrating \eqref{eq:proof-prop-dynamic-triggering-lyap-noise-free} along the solutions to (\ref{eq:closed-loop}), (\ref{eq:e}), (\ref{eq:eta-noise-free}), (\ref{eq:triggering-dynamic-noise-free}), noting that $U(x,\eta)$ is not affected by jumps at the triggering instants, and exploiting the expression of $U$, we deduce that item (b) of Proposition \ref{prop:dynamic-triggering-rule-noise-free} holds. \hfill $\blacksquare$

According to Proposition \ref{prop:dynamic-triggering-rule-noise-free}, given a matrix $\widetilde\Psi$, which satisfies \eqref{eq:SDP_triggering-quadratic}, we can then design a dynamic version of the triggering condition in Section \ref{subsect:quadratic-trig} in the noise-free case. The convergence of the $x$-component of the solutions is still exponential as the time goes to infinity in view of item (b) of Proposition \ref{prop:dynamic-triggering-rule-noise-free}.

In the noisy case, the dynamics of variable $\eta$ needs to be modified as,  similarly to \cite{Dolk-et-al-tac17},
\begin{equation}
\begin{array}{rlll}    
\dot \eta(t)    \in  -\lambda \eta(t) - \psi(t-t_k)\left(z(t)^{\top}\widetilde\Psi z(t)-\nu\right),
\end{array}\label{eq:eta-noisy}
\end{equation}
with $\lambda>0$ arbitrary and $\widetilde\Psi$ symmetric to be designed. Given $\overline\tau_d(\sigma_1)$ as in Section \ref{subsect:quadratic-trig},  namely $\overline\tau_d(\sigma_1)\in \{0, \tau_d(\sigma_1)\}$, $\tau_d(\sigma_1)$ as in in \eqref{eq:tau_d} and  $\sigma_1$ as in \eqref{eq:sigma-time-reg}, $\psi$ is defined as follows. When $\overline{\tau}_d(\sigma_1)=\tau_d(\sigma_1)$, $\psi(s)=0$ for $s\in[0,\overline{\tau}_d(\sigma_1))$, $\psi(s)=1$ for $s>\overline{\tau}_d(\sigma_1)$ and $\psi(s)=[0,1]$ for $s=\overline{\tau}_d(\sigma_1)$. When $\overline{\tau}_d(\sigma_1)=0$, $\psi(s)=1$ for any $s\geq 0$.  In \eqref{eq:eta-noisy}, the  term  $z(t)^{\top}\widetilde\Psi z(t)-\nu$, which is related to  \eqref{eq:triggering-quadratic-noisy}, is filtered after $\overline\tau_d(\sigma_1)$ units of time have elapsed. We note that (\ref{eq:eta-noisy}) is a differential \emph{inclusion} when $\overline\tau_d(\sigma)=\tau_d(\sigma)$ because $\psi$ is multi-valued at $\tau_d(\sigma)$, and solutions to (\ref{eq:eta-noisy}) are understood in the Krasovskii sense on $[t_k,t_{k+1}]$ with $k\in\mathcal{N}$ in this case, see \cite[Chapter 4.5]{Goebel-Sanfelice-Teel-book}. The triggering rule  becomes: $t_0=0$ and 
{\setlength\arraycolsep{2pt} 
\begin{equation} \label{eq:triggering-dynamic-noisy}
\begin{array}{rlll}
t_{k+1} & = & 
\inf \Big\{t \in\R\,:\,t\geq t_k+\overline{\tau}_d(\sigma_1) \\
& &\hspace{1.2cm} \text{and } \eta(t)-\theta(z(t)^\top\widetilde{\Psi}z(t) - \nu) \leq 0  \Big\},
\end{array}
\end{equation}}%
with arbitrary $\theta\geq 0$ and, again, $\sigma_1$ as in \eqref{eq:sigma-time-reg}. We have the next result.

\begin{proposition}\label{prop:dynamic-triggering-rule-noisy} Suppose the conditions of Proposition \ref{prop:quadratic-triggering-rule-noisy} are satisfied. If $\overline{\tau}_d(\sigma_1)=\tau_d(\sigma_1)$ or $\nu>0$, system (\ref{eq:closed-loop2}), (\ref{eq:e-noisy}), (\ref{eq:eta-noisy}), (\ref{eq:triggering-dynamic-noisy}):
\begin{enumerate}
    \item[(a)] admits global inter-event time $\max\{\overline{\tau}_d(\sigma_1),\widetilde\tau(\sigma_2)\}$ as in item (a) of Proposition \ref{prop:quadratic-triggering-rule-noisy};
    \item[(b)] is (practically) ISS, in particular there exist $c_1\geq 1$ and $c_2,c_3,c_4>0$ such that any solution  $(x,\eta)$ with disturbance input $d$ satisfies  $|(x(t),\sqrt{\eta(t)})|\leq c_1 e^{-c_2 t}|(x(0),\sqrt{\eta(0)})|+c_3\|d\|_{[0,t]}+c_4\nu$. \hfill $\Box$
\end{enumerate}
\end{proposition}

\emph{Proof.} Like in the proof of Proposition \ref{prop:dynamic-triggering-rule-noise-free}, we have that $\eta\geq 0$ along the solutions to  (\ref{eq:closed-loop2}), (\ref{eq:e-noisy}), (\ref{eq:eta-noisy}), (\ref{eq:triggering-dynamic-noisy}). Moreover, for any solution $(x,\eta)$ to (\ref{eq:closed-loop2}), (\ref{eq:e-noisy}), (\ref{eq:eta-noisy}), (\ref{eq:triggering-dynamic-noisy})  with input disturbance $d$, at any  $k\in\mathcal{N}$, the next inter-transmission time is greater than the corresponding inter-transmission time generated by (\ref{eq:triggering-quadratic-noisy}) at the same state. We then invoke item (a) of Proposition \ref{prop:quadratic-triggering-rule-noisy} to derive that item (a) of Proposition \ref{prop:dynamic-triggering-rule-noisy} holds.

To prove item (b) of Proposition \ref{prop:dynamic-triggering-rule-noisy}, we consider the same Lyapunov function as in the proof of Proposition \ref{prop:dynamic-triggering-rule-noise-free}, namely  $U(x,\eta)=V(x)+\eta/\mu$ with $V(x)=x^\top S x$ for any $z=(x,e)\in\R^{2n}$ and $\eta\in\R_{\geq 0}$ with $\mu$ coming from (\ref{eq:SDP_triggering-quadratic-noisy}). Let $(x,\eta)$ be a solution to (\ref{eq:closed-loop2}), (\ref{eq:e-noisy}), (\ref{eq:eta-noisy}), (\ref{eq:triggering-dynamic-noisy}) with input disturbance $d$, and $k\in\mathcal{N}$. We treat  the case where $\overline\tau_d(\sigma_1)=\tau_d(\sigma_1)$; the proof follows similar developments when $\overline\tau_d(\sigma_1)=0$. 

When $t\in[t_k,t_k+\tau_d(\sigma_1))$, we have from the definition of $\tau_d(\sigma_1)$ that (\ref{eq:bound_e}) holds by Lemma \ref{lem:intersampling_ISS_data}. Hence, in view of (\ref{eq:proof-prop-quad-noisy-lyap-ineq-before-taud}) and \eqref{eq:eta-noisy}, 
(\ref{eq:proof-prop-quad-noisy-lyap-ineq-before-taud}) 
{\setlength\arraycolsep{2pt} 
\begin{equation}
\begin{array}{rlll}
\dot U(x(t),\eta(t)) & \leq  & -(\omega_1-\sigma_1\omega_2)|x(t)|^2\\
& & +(\sigma_1\omega_2+\omega_3)|x(t)| \|d\|_{[0,t]} - \lambda/\mu\, \eta(t),  
\end{array}\label{eq:prop-dyn-etc-noisy-lyap-before-tau-d}
\end{equation}}%
as in this case $\psi(t-t_k)=0$ and  $\omega_1-\sigma_1\omega_2>0$ for  (\ref{eq:sigma-time-reg}) holds.

When $t\in(t_k+\tau_d(\sigma_1),t_{k+1})$, we have from (\ref{eq:V_approx_ISS}), 
{\setlength\arraycolsep{2pt} 
\begin{equation}\nonumber
\begin{array}{rlll}
\dot U(x(t),\eta(t)) & \leq  & - x(t)^\top \left( \frac{S \Omega S}{2} \right) x(t) + z(t)^\top\overline M(D_0)z(t)\\
& & + 2 d(t)^\top S x(t)  - \lambda/\mu\, \eta(t)  \\
& & - 1/\mu\,z(t)^{\top}\widetilde \Psi z(t) + \nu/\mu.
\end{array}
\end{equation}}%
We then follow similar developments as in the proof of Proposition \ref{prop:quadratic-triggering-rule-noisy} to conclude that 
{\setlength\arraycolsep{2pt} 
\begin{equation}
\begin{array}{rlll}
\dot U(x(t),\eta(t)) & \leq  & - x(t)^\top \left( \frac{S \Omega S}{2} \right) x(t) \\
& & + 2 d(t)^\top S x(t)  - \lambda/\mu\, \eta(t)   + \nu/\mu.
\end{array}\label{eq:prop-dyn-etc-noisy-lyap-after-tau-d}
\end{equation}}%

When $t=t_k+\tau_d(\sigma)$,
{\setlength\arraycolsep{2pt} 
\begin{equation}
\begin{array}{rlll}\label{eq:prop-dyn-etc-noisy-lyap-at-tau-d-prel}
\dot U(x(t),\eta(t)) & = & \dot V(x(t)) - \lambda/\mu\, \eta(t) \\
& &- c_{\psi}/\mu\,(z(t)^\top\widetilde\Psi z(t)-\nu),
\end{array}
\end{equation}}%
with $c_{\psi}\in[0,1]$. If $z(t)^{\top}\widetilde \Psi z(t) \geq 0$, 
$\dot U(x(t),\eta(t))  =  \dot V(x(t)) - \lambda/\mu\, \eta(t) + c_{\psi}/\mu\,\nu
\leq  \dot V(x(t)) - \lambda/\mu\, \eta(t) + \nu/\mu$, 
and we invoke \eqref{eq:V_approx_ISS} and (\ref{eq:bound_e}) to derive that
{\setlength\arraycolsep{2pt} 
\begin{equation}
\begin{array}{rlll}
\dot U(x(t),\eta(t)) & = & - (\omega_1 - \sigma_1\omega_2) |x(t)|^2
\\ 
& & 
+ (\sigma_1\omega_2 +  \omega_3 ) |x(t)| \|d\|_{[0,t]} \\
& & - \lambda/\mu\, \eta(t)  + \nu/\mu.
\end{array}\label{eq:prop-dyn-etc-noisy-lyap-at-tau-d}
\end{equation}}%
If $z(t)^{\top}\widetilde \Psi z(t) \leq 0$, we deduce from (\ref{eq:prop-dyn-etc-noisy-lyap-at-tau-d-prel}), 
{\setlength\arraycolsep{2pt} 
\begin{equation}
\begin{array}{rlll}
\dot U(x(t),\eta(t)) & = & \dot V(x(t)) - \lambda/\mu\, \eta(t) \\
& & - 1/\mu\,z(t)^\top\widetilde\Psi z(t) + \nu/\mu,
\end{array}
\end{equation}}%
and we proceed as in the case where $t\in(t_k+\tau_d(\sigma_1),t_{k+1})$ to derive (\ref{eq:prop-dyn-etc-noisy-lyap-after-tau-d}). Item (b) of Proposition \ref{prop:dynamic-triggering-rule-noisy}  follows by integration from (\ref{eq:prop-dyn-etc-noisy-lyap-before-tau-d}), (\ref{eq:prop-dyn-etc-noisy-lyap-after-tau-d}) and (\ref{eq:prop-dyn-etc-noisy-lyap-at-tau-d}), noting that $U(x,\eta)$ is not affected by jumps at the triggering instants. 
\hfill $\blacksquare$

\begin{rem}\label{rem:qi-et-al} Compared to \cite{qi-et-al-tie2022}, the triggering condition in \eqref{eq:triggering-dynamic-noisy} is more general as: (i) it involves a general quadratic term; (ii)  $\theta$ can be non-zero: (iii) a fixed threshold-like parameter $\nu$ is allowed, which may help reducing the number of transmissions at the price of a practical ISS property. \hfill $\Box$
\end{rem}

\subsection{Decreasing threshold on the Lyapunov function}\label{subsect:decreasing-V}

We present a last triggering rule, which consists in imposing  a given (decreasing) threshold on the Lyapunov function $V$ evaluated along the solutions to \eqref{system}, \eqref{control}, as suggested in e.g., \cite{Wang-Lemmon-aut11,Mazo-Anta-Tabuada-Aut10} in  model-based settings. Exploiting the fact that the Lyapunov function $V$ can be derived from data as shown in Sections \ref{sec:exact} and \ref{sec:noise}, we present ``robustified'' data-based versions of these sampling strategies.

In the noise-free case and in absence of network, we have that $V(x)=x^\top S x$ is a Lyapunov function for $\dot x(t)=(A+BK)x(t)$ where $S=(X_0Y)^{-1}$, $Y$ is a solution to \eqref{eq:SDP} and $K=U_0 Y(X_0 Y)^{-1}$, see Section \ref{subsect:trig-rule-noisefree}. We then have $\left\langle \nabla V(x),(A+BK)x\right\rangle \leq -\rho_1 V(x)$ 
for any $x\in\R^{n}$,
where  $\rho_1>0$ is any (sufficiently small) constant such that $-\rho_1 S \succeq (A+BK)^\top S + S(A+BK)=(X_1 G)^\top S + S(X_1 G)$. Note that the condition $-\rho_1 S \succeq (X_1 G)^\top S + S(X_1 G)$ is data-based and can thus be exploited to find $\rho_1$ using set $\mathbb{D}$ in \eqref{dataset}. Based on this observation, we introduce the threshold variable
$\eta\in\R_{\geq 0}$, whose dynamics is 
\begin{equation}
\begin{array}{rllll}
\dot{\eta}(t) = - \varsigma\rho_1 \eta(t), & \eta(0)\geq V(x(0)),
\end{array}\label{eq:eta-decay-V-noise-free}
\end{equation}
where $\varsigma\in(0,1)$ is arbitrary. Hence $\eta$ upper-bounds $V(x)$ along the solutions to the closed-loop system $\dot x(t)=(A+BK)x(t)$, as $\eta(0)\geq V(x(0))$ and $\dot{\eta}(t)\geq \dot{V}(x(t))$ in view of the comparison principle \cite[Chapter 3.4]{KHALIL2002}. The triggering rule is then given by: $t_0=0$ and\footnote{Like in footnote \ref{footnote:dynamic-etc-theta=0}, triggering occurs when $V(x(t))=\eta(t)$ \emph{and} $\dot{V}(x(t))\geq \dot{\eta}(t)$, which we do not specify consistently with related model-based results of the literature and not to overload \eqref{eq:triggering-decay-V}. This extra condition will not be required in the noisy case in \eqref{eq:triggering-decay-V-noisy} thanks to time-regularization.}
\begin{equation} \label{eq:triggering-decay-V}
t_{k+1} = \left\{ 
\begin{array}{l}
\inf \left\{t \in\R\,:\,t> t_k \text{ and } V(x(t))= \eta(t) \right\}   \\[0.1cm]
\hfill \text{if } (x(t_k),\eta(t_k))\neq0,  \\[0.2cm]
 + \infty \hfill \text{otherwise.}
\end{array}
\right.
\end{equation}
We have the next result in the noise-free case.

\begin{proposition} \label{prop:decay-V-noise-free} Suppose Assumption \ref{ass:rich} holds. Let  $Y$ be any solution to \eqref{eq:SDP}, $S=(X_0Y)^{-1}$, $K=U_0Y(X_0Y)^{-1}$
and let $\rho_1>0$  be such that $(X_1 G)^\top S + S(X_1 G) \preceq -\rho_1 S$. Furthermore, let
$L$ be as in \eqref{eq:L}, $\varsigma\in(0,1)$ arbitrary, and $\mu,\,\sigma>0$  be such that the following SDP (in the decision variables
$\mu$ and
$\sigma^2$) is satisfied
\begin{equation}
\label{eq:SDP_triggering_varsigma}
\mu M_ \varsigma- \Psi(\sigma) \preceq 0
\end{equation}
where the matrix $M_\varsigma$ is the same as matrix $M$ in
\eqref{eq:V_exact} with $-Q$ replaced by $-Q+\varsigma \rho_1 S$ and $\Psi(\sigma)$ is as in \eqref{eq:Psi}.
Let $V(x)=x^{\top}Sx$, then  system \eqref{eq:closed-loop}, \eqref{eq:e}, \eqref{eq:eta-decay-V-noise-free}, \eqref{eq:triggering-decay-V}:
\begin{itemize}
\item[(a)] admits $\tau(\sigma)$  in item (a) of Theorem \ref{thm:exact_sampling} as global minimum inter-event time with $\sigma$ ensuring \eqref{eq:SDP_triggering_varsigma};
\item[(b)] is  globally asymptotically stable, in particular there exist $c_1\geq 1$ and $c_2>0$ such that any solution  $(x,\eta)$ satisfies $|(x(t),\!\sqrt{\eta(t)})|\leq c_1 e^{-c_2 t}|(x(0),\!\sqrt{\eta(0)})|$ for any $t\!\geq\! 0$.
\hfill $\Box$
\end{itemize}
\end{proposition}

\emph{Proof.} 
We first show the feasibility of \eqref{eq:SDP_triggering_varsigma}. Let $\sigma:=\sqrt{\mu/c}$
with $c>0$ any sufficiently large constant such that $-Q +\varsigma \rho_1 S+ I/c \prec 0$,
where $Q$ is defined in (\ref{eq:Q}); clearly, this $c$ exists because $Q -\varsigma\rho_1 S\succ 0$ (as $Q\succeq \rho_1 S$ and $\varsigma\in(0,1)$).
For such a choice of $\sigma$, we have $\mu(- Q + \varsigma\rho_1 S)+\sigma^2 I \prec 0$
for every $\mu>0$. Therefore, for this choice of $\sigma$, \eqref{eq:SDP_triggering_varsigma}
is equivalent by a Schur complement to the two conditions
$\mu (-Q+ \varsigma\rho_1 S+I/c) \prec 0$  and 
$-I + \mu (SX_1L)^\top (Q - \varsigma\rho_1 S - I/c)^{-1} (SX_1L) \prec 0$, 
which are jointly satisfied for $\mu$ sufficiently small.


In view of (\ref{eq:SDP_triggering_varsigma}), for any $z\in\R^{2n}$, $z^{\top} \Psi(\sigma) z< 0$ implies  $\left\langle \nabla V(x),(A+BK)x+BKe\right\rangle=z^\top M z < -\varsigma\rho_1 x^\top S x=-\varsigma\rho_1 V(x)$. This implies that for any solution $x$ to \eqref{eq:closed-loop}, \eqref{eq:e}, \eqref{eq:eta-decay-V-noise-free}, \eqref{eq:triggering-decay-V}, for any $t\in[t_k,t_{k+1}]$, as long as  $z(t)^\top \Psi(\sigma)z(t)<0$ (which is the case right after a transmission whenever $x(0)\neq 0$), we have $\dot{V}(x(t))< -\varsigma\rho_1 V(x(t))$ and thus $V(x(t))< \eta(t)$ in view of \eqref{eq:eta-decay-V-noise-free} and \eqref{eq:triggering-decay-V}. Hence, as the time it takes for $z^\top \Psi(\sigma)z$ to grow from $-\sigma^2|x|$ to $0$ along \eqref{eq:closed-loop}, \eqref{eq:e} is lower bounded by $\tau(\sigma)$ in view of item (a) of Theorem \ref{thm:exact_sampling}, the time it takes for $V(x)$ to reach $\eta$, which is the inter-transmission time generated by \eqref{eq:triggering-decay-V}, is bigger than the former and is thus  also lower bounded by $\tau(\sigma)$. Item (a) of Proposition \ref{prop:decay-V-noise-free} is then satisfied as when $x(0)=0$ no transmissions ever occur. 


The proof of item (b) follows by considering Lyapunov function $U(x,\eta)=\max\{V(x),\eta\}$ for any $x\in\R^n$ and $\eta\geq 0$, and exploiting the fact that $\eta$ upper-bounds $V$ along the solutions to (\ref{eq:closed-loop}), (\ref{eq:e}), (\ref{eq:eta-decay-V-noise-free}), (\ref{eq:triggering-decay-V}) and strictly decreases outside the origin. \hfill $\blacksquare$

We notice that the same type of stability property is stated in items (b) of Propositions \ref{prop:dynamic-triggering-rule-noise-free} and  \ref{prop:decay-V-noise-free}. 

In the noisy case, the ISS Lyapunov function is given by $V(x)=x^\top S x$ for any $x\in\R^{n}$ with $S=(X_0 Y)^{-1}$ and $Y$ a solution to \eqref{eq:2SDP}, assuming it exists, see Section \ref{subsect:triggering-rule-noisy}. We modify the dynamics of threshold $\eta$ in (\ref{eq:eta-decay-V-noise-free}) as 
\begin{equation}
\begin{array}{rllll}
\dot{\eta}(t)  =  - \varsigma_d \eta(t) + \nu, & \eta(0)\geq V(x(0)) 
\end{array}\label{eq:eta-decay-V-noisy}
\end{equation}
where $\varsigma_d>0$ and $\nu\geq 0$ are both arbitrary. We emphasize that the decay rate of $\eta$, namely $\varsigma_d$, can take any value in $\R_{>0}$, which was not the case in (\ref{eq:eta-decay-V-noise-free}); we elaborate more on this point in Remark \ref{rem-decay-V-rate} below. 
We then ``robustify'' the triggering rule (\ref{eq:triggering-decay-V}) as: $t_0=0$ and  
\begin{equation} \label{eq:triggering-decay-V-noisy}
t_{k+1} \!=\! 
\inf \left\{t \in\R\,\!:\,\!t\geq t_k\!+\!\tau_d(\sigma) \text{ and } V(x(t))\!\geq\!  \eta(t) \right\}, 
\end{equation} 
with $\sigma$ and $\tau_d(\sigma)$ as in (\ref{eq:sigma-time-reg}) and (\ref{eq:tau_d}), respectively. 
We consider $\tau_d(\sigma)$ in (\ref{eq:triggering-decay-V-noisy}) and not $\overline\tau_d(\sigma)$ as in the other techniques covered so far in this section to simplify the exposition; otherwise the selection of $\nu$  would need to depend on some known upper-bound on  $\|d\|_\infty$. In view of \eqref{eq:triggering-decay-V-noisy},  the existence of a global minimum inter-event time for system (\ref{eq:closed-loop2}), (\ref{eq:e-noisy}), (\ref{eq:eta-decay-V-noisy}), (\ref{eq:triggering-decay-V-noisy}) is immediate and we have the next stability result.

\begin{proposition}\label{prop:decay-V-noisy} Suppose the following holds. 
\begin{enumerate}
    \item[(i)] Assumptions \ref{ass:rich} and \ref{ass:D0} are verified with $\Delta$ given.
    \item[(ii)] Let $\Omega \succ 0$, SDP \eqref{eq:2SDP} is feasible and $K=U_0Y(X_0Y)^{-1}$ is the resulting controller as in Theorem \ref{thm:approx}.
\end{enumerate}
Let $V(x)=x^{\top}Sx$ for any $x\in\R^{n}$ with $S=(X_0 Y)^{-1}$ and $Y$ from (\ref{eq:2SDP}), then system (\ref{eq:closed-loop2}), (\ref{eq:e-noisy}), (\ref{eq:eta-decay-V-noisy}), (\ref{eq:triggering-decay-V-noisy}) is (practically) exponentially ISS, in particular there exist $c_1\geq 1$ and $c_2,c_3,c_4>0$ such that any solution  $(x,\eta)$ with disturbance input $d$, it holds that $|(x(t),\sqrt{\eta(t)})|\leq c_1 e^{-c_2 t}|(x(0),\sqrt{\eta(0)})|+c_3\|d\|_{[0,t]}+c_4\nu$.\hfill $\Box$
\end{proposition}

\emph{Proof.} 
We consider the same Lyapunov function as in the proof of Proposition \ref{prop:decay-V-noise-free}, namely $U(x,\eta)=\max\{V(x),\eta\}$ for any $x\in\R^n$ and $\eta\geq 0$. Let $(x,\eta)$ be a solution to (\ref{eq:closed-loop2}), (\ref{eq:e-noisy}), (\ref{eq:eta-decay-V-noisy}), (\ref{eq:triggering-decay-V-noisy}) with input disturbance $d$. The existence of such a solution is ensured as there exists  a minimum inter-event time $\overline\tau_d(\sigma)$, thereby excluding Zeno phenomenon, and the  involved dynamics is linear. Let $k\in\mathcal{N}$ and $t\in[t_k,t_{k+1}]$. To differentiate $U$ along $(x,\eta)$ requires care as $U$ is not differentiable everywhere, but almost everywhere. A convenient tool in this context is Clarke's  derivative \cite{clarke1990optimization}. To avoid introducing too many technicalities, we exploit \cite[Lemma 1]{Liberzon-Nesic-Teel-cdc12}, which states in our case that the Clarke's  derivative of $U(x(t),\eta(t))$ is $\dot\eta(t)$ when 
$\eta(t)>V(x(t))$, it is $\dot{V}(x(t))$ when 
$\eta(t)<V(x(t))$, and it is less than or equal to $\max\{\dot{V}(x(t)),\dot\eta(t)\}$ when $V(x(t))=\eta(t)$. By deriving  ISS dissipation inequalities for $\dot{V}(x(t))$ and $\dot{\eta}(t)$, we then obtain the desired result by invoking \cite[p.99]{Teel-Praly-mcss-00} and integrating the obtained inequalities as in the other proofs of this work. We now analyse  $\dot{V}(x(t))$ and $\dot{\eta}(t)$. 

We have that the $\eta$-system satisfies a suitable ISS dissipation inequality by design in view of (\ref{eq:eta-decay-V-noisy}). Regarding $\dot{V}(x(t))$, if $t\in[t_k,t_{k}+\tau_d(\sigma)]$, we proceed as before in the paper and exploit the fact that, by definition of $\tau_d(\sigma)$ and since $\sigma$ satisfies \eqref{eq:sigma-time-reg},  \eqref{eq:proof-prop-quad-noisy-lyap-ineq-before-taud} holds. If $t\in[t_k+\tau_d(\sigma),t_{k+1})$, then $V(x(t))<\eta(t)$, $U(x(t),\eta(t))=\eta(t)$ and the Clarke's derivative is given by $\dot{\eta}(t)=-\varsigma_d\,\eta(t)+\nu=-\varsigma_d\, U(x(t),\eta(t))+\nu$.  Based on these properties, and since $U$ is not affected by jumps at triggering instants along the solutions, we derive the stability property stated in  Proposition \ref{prop:decay-V-noisy}. \hfill $\blacksquare$

\begin{rem}\label{rem-decay-V-rate} It is because \eqref{eq:triggering-decay-V-noisy} is based on time-regularization that $\varsigma_d$ can take any value in $\R_{>0}$ in \eqref{eq:eta-decay-V-noisy}. When $\varsigma_d$  is bigger than the decay rate of the Lyapunov function $V$ along the solutions to the considered system, a transmission instant will occur immediately at $t_k+\tau_d(\sigma)$ with $k\in\mathcal{N}$ and $\tau_d(\sigma)$-periodic sampling will occur. A reasonable option is to select $\varsigma_d>0$ such that $\varsigma_d S \prec S\Omega S$ in view of (\ref{eq:V_approx}c), so that the decay rate of $\eta$ is not faster than the one of $V(x)$ along the solutions to the system ignoring sampling and $d$. \hfill $\Box$
\end{rem}




\section{Concluding remarks} \label{sec:conc}

We have presented an approach to design robust event-triggered state-feedback controllers for unknown stabilizable perturbed linear time invariant systems directly based on a collection of noisy off-line data and not a model of the dynamics. In particular, we have derived data-based version of the event-triggered control strategies originally developed in a model-based settings in \cite{Tabuada07,Girard-tac15,Dolk-et-al-tac17,wang-et-al-cdc2020(small-gain),Borgers-Heemels-tac14,Mazo-Anta-Tabuada-Aut10}. 

The tools developed in this work may provide foundations to develop data-based event-triggered control designs for other control scenarios including when the (continuous-time) plant has nonlinear dynamics, when only an output, and not the full state as in this paper, is available, or when other than zero-order-hold holding strategies are used to implement the control input, to mention a few promising research directions. 

\appendix 

\subsection{Alternative data acquisition scheme} \label{app:int}

The results presented in this work rest on $X_1$ in (\ref{eq:data3}). The computation of $X_1$ can be error-prone
as it involves computing the derivative of $x$. We briefly discuss a
data collection scheme that can be used when the derivative of $x$
is difficult to compute. The idea is to consider the 
\emph{integral} version of the relation 
$\dot x = A x + B u+d$, for any $\tau_1\leq \tau_2$,
$\int_{\tau_1}^{\tau_2} \dot x(t) \mathrm{d}t = 
\int_{\tau_1}^{\tau_2} (A x(t) + B u(t) + d(t)) \mathrm{d}t$, 
which writes equivalently as
$x(\tau_2) - x(\tau_1) = 
\displaystyle A \int_{\tau_1}^{\tau_2} x(t) \mathrm{d}t
+ B \int_{\tau_1}^{\tau_2} u(t) \mathrm{d}t + \int_{\tau_1}^{\tau_2} d(t) \mathrm{d}t$. 
We can therefore choose a sampling time $T_s>0$ (this is just
for simplicity, we do not need periodicity) to obtain
\begin{equation}
\label{eq:app_b_3}
\begin{array}{l}
\underbrace{x((k+1)T_s) - x(kT_s)}_{=:\xi(k)} = \\[.7cm]
A \underbrace{\int_{kT_s}^{(k+1)T_s} x(t) \mathrm{d}t}_{=:r(k)} + 
B \underbrace{\int_{kT_s}^{(k+1)T_s} u(t) \mathrm{d}t}_{=:v(k)} + 
\underbrace{\int_{kT_s}^{(k+1)T_s} d(t) \mathrm{d}t}_{=:w(k)}
\end{array}
\end{equation}
with $k \geq 0$. This represents the relation for the $k$-th sample.
Let $T>0$ be the number of samples that we want to collect. 
If we define $\underline X_1 := \begin{bmatrix} \xi(0) 
& \xi(1) & \ldots & \xi(T-1) \end{bmatrix}$,  $\underline X_0 := \begin{bmatrix} r(0) & r(1) & \ldots & r(T-1) \end{bmatrix}$, $\underline U_0 := \begin{bmatrix} v(0) & v(1) & 
\ldots & v(T-1) \end{bmatrix}$, $\underline D_0 := \begin{bmatrix} w(0) & w(1) & \ldots & w(T-1) \end{bmatrix}$, 
we get 
$\underline X_1=A\underline X_0+B\underline U_0+\underline D_0$, which is the integral version of the relation
$X_1=AX_0+BU_0+D_0$ considered in this work. 
We can therefore restate all the results with $U_0,X_0,X_1,D_0$ replaced
by $\underline U_0,\underline X_0,\underline X_1,\underline D_0$.
Different from $X_1$, $\underline X_1$ does not involve 
the derivative of $x$. 
  
\subsection{Proof of Lemma \ref{lem:Petersen}} \label{sec:petersen}
 
Lemma \ref{lem:Petersen} is a direct consequence of the next result. 

\begin{lem} \label{lem:Petersen_aux}
Let $B \in \mathbb R^{n \times p}$ and $C \in \mathbb R^{q \times n}$ be given matrices. Then 
for any $\epsilon > 0$ and any $D\in\mathcal{D}$ with $\mathcal{D}$ in \eqref{eq:noise_model}, we have  
$BD^\top C + C^\top D B^\top \preceq 
\epsilon^{-1} B B^\top + \epsilon C^\top \Delta \Delta^\top C$. \hfill $\Box$
\end{lem} 
\emph{Proof}. A completion of squares
$( \sqrt{\epsilon^{-1}} B -
\sqrt{\epsilon} C^\top D ) 
( \sqrt{\epsilon^{-1}} B - 
\sqrt{\epsilon} C^\top D )^\top
\succeq 0$ 
gives the result. \hfill $\blacksquare$

\emph{Proof of Lemma \ref{lem:Petersen}.}
Let \eqref{eq:2SDP1} hold. By a Schur complement, this is equivalent to 
$X_1Y + (X_1Y)^\top + \Omega +
\epsilon^{-1} Y^\top Y + \epsilon  \Delta \Delta^\top \prec 0$.  
By applying Lemma \ref{lem:Petersen_aux} with $B=-Y^\top$ and $C=I$, we obtain
$X_1Y + (X_1Y)^\top + \Omega - D Y - (D Y)^\top \prec 0 \quad \forall D \in \mathcal D$, 
which corresponds to \eqref{eq:2SDP_noisy}. \hfill $\blacksquare$

\bibliographystyle{IEEEtran}

\bibliography{refs}

\end{document}